\newif\iflncs\lncsfalse
\newif\ifmai\maifalse
\newif\ifmai\maifalse
\newif\iffull\fullfalse
\documentclass[submission,copyright,creativecommons]{eptcs}
\providecommand{\event}{GandALF 2025} % Name of the event you are submitting to

\usepackage{iftex}

\ifpdf
  \usepackage{underscore}         % Only needed if you use pdflatex.
  \usepackage[T1]{fontenc}        % Recommended with pdflatex
\else
  \usepackage{breakurl}           % Not needed if you use pdflatex only.
\fi

\usepackage[T1]{fontenc}
\usepackage{wrapfig}
\usepackage{graphicx}
\usepackage{subcaption}
\usepackage[T1]{fontenc}
\usepackage[utf8]{inputenc}
\usepackage{listings}
\usepackage[figuresright]{rotating}
\usepackage{courier}
\usepackage{float}
\usepackage{algorithm}
\usepackage{algpseudocode}
\usepackage{graphicx}
\usepackage{algorithm}
\usepackage{tikz}
\usetikzlibrary{decorations.pathreplacing,calc}
\newcommand{\tikzmark}[1]{\tikz[overlay,remember picture] \node (#1) {};}
\usepackage{algpseudocode}
\usepackage{listings}
\usepackage{ifthen}
\usepackage{verbatimbox}
\usepackage{svg}
\usepackage{chngcntr}
\usepackage{apptools}
\usepackage{booktabs}
\usepackage{multirow}
\usepackage{graphicx}
\usepackage{pgf}
\usepackage{tikz}
\usepackage{wrapfig}
\usepackage{subcaption}
\usepackage{url}
\usepackage{listings}
\usepackage{moresize}
\usepackage{lipsum}
\usepackage{algorithm}
\usepackage{algpseudocode}
\usepackage{tikz}

\newcommand*{\AddNote}[4]{%
	\begin{tikzpicture}[overlay, remember picture]
		\draw [decoration={brace,amplitude=0.5em},decorate,very thick]
		($(#3)!(#1.north)!($(#3)-(0,1)$)$) --
		($(#3)!(#2.south)!($(#3)-(0,1)$)$)
		node [align=center,  pos=0.5, anchor=west] {#4}; %text width=2.5cm,
	\end{tikzpicture}
}%

\newcommand{\lui}[1]{}%{\marginpar{ \textbf{Nota:} {\footnotesize #1}}}
\newcommand{\reals}{{\mathbb{R}}}

\newcommand{\T}{\ensuremath{\Theta}}

\newcommand{\supp}{\mathrm{supp}}
\newcommand{\es}{\emptyset}
\usepackage{color}
\newcommand\qedsymbol{\textcolor{gray}{\ensuremath{\blacktriangleleft}}}
\newenvironment{proofof}[1]{ {\it\noindent Proof of }%{\sc\noindent Proof of }
{#1}.}{ \hfill \qedsymbol \vskip .3mm}

\newcommand{\X}{\mathcal{X}}

\newcommand{\St}{\mathcal{P}}

\newcommand{\F}{\mathcal{F}}

\newcommand{\sem}[1]{[\![#1]\!]}
\newcommand{\ereals}{\overline\reals}
\newcommand{\erealspl}{\overline\reals^+}
\newcommand{\expc}{\mathrm{E}}
\newcommand{\extF}{{\F}}

\newcommand{\PP}{\mathcal{P}}
\newcommand{\K}{\kappa}

\newcommand{\nil}{\text{\rm\textsf{nil}}}
\newcommand{\ppg}{\mathbf{G}}
\newcommand{\psco}{\mbox{$\mathbf{sc}$}}
\newcommand{\iomega}{\tilde\omega}
\newcommand{\Cyl}{\mathcal{C}}
\newcommand{\cy}[1]{\mathfrak{c}(#1)}%{\check{#1}}

\newcommand{\term}{\text{\rm\textsf{T}}}

\newcommand{\scoz}{\mathrm{sc}}
\newcommand{\wt}{\mathrm{w}}
\newcommand{\lift}[1]{\check{#1}}
\newcommand{\FC}{\mathrm{FK}}
\newcommand{\mufk}{\phi}
\newcommand{\TSIpf}{VPF}
\newcommand{\vsp}{\vspace*{.2cm}}
\newcommand{\mik}[1]{\marginpar{ \textbf{Mic:} {\footnotesize #1}}}

\def\th@plain{%
\thm@notefont{}% same as heading font
\itshape % body font
}
\def\th@definition{%
\thm@notefont{}% same as heading font
\normalfont % body font
}
\makeatother

\iflncs{}\else \newtheorem{definition}{Definition}\fi

\iflncs{}\else \fi

\iflncs{}\else \newtheorem{theorem}{Theorem}\fi

\iflncs{}\else  \fi

\iflncs{}\else \newtheorem{lemma}{Lemma}\fi

\iflncs{}\else \newtheorem{remark}{Remark}\fi
\iflncs{}\else \newtheorem{example}{Example}\fi

\usepackage{array}
\newcommand{\fini}{\mathrm{f}}
\newcommand{\pr}{\mathrm{pr}}
\usepackage{chngcntr}
\usepackage{apptools}
\AtAppendix{\counterwithin{lemma}{section}}
\AtAppendix{\counterwithin{theorem}{section}}
\AtAppendix{\counterwithin{corollary}{section}}
\AtAppendix{\counterwithin{proposition}{section}}
\AtAppendix{\counterwithin{theorem}{section}}
\AtAppendix{\counterwithin{definition}{section}}
\newenvironment{proof}{\noindent{\sc Proof.}~}{ \hfill $\Box$ \vskip 2mm }
\usepackage{amssymb,amsmath}
\title{Parallelizable Feynman-Kac Models\\
	for Universal Probabilistic Programming\thanks{Work partially supported by the project SERICS (PE00000014), EU funded  NRRP MUR program  - NextGenerationEU.}}
\author{
	Michele Boreale \qquad\qquad Luisa Collodi
\institute{Università degli Studi di Firenze, Dipartimento di Statistica, Informatica, Applicazioni “G. Parenti”,}
	\email{\{michele.boreale,luisa.collodi\}@unifi.it}
}

\def\titlerunning{Parallelizable Feynman-Kac Models
	for Universal Probabilistic Programming}
\def\authorrunning{M. Boreale, L. Collodi}
\begin{document}
\maketitle

\begin{abstract}
We study   provably correct and efficient instantiations of  Sequential Monte Carlo (SMC) inference  in the context of formal operational semantics of Probabilistic Programs (PPs). We focus on \emph{universal} PPs   featuring  sampling from arbitrary measures and conditioning/reweighting in unbounded loops. We first equip Probabilistic Program Graphs (PPGs), an automata-theoretic description format of PPs, with an expectation-based semantics over infinite execution traces, which also incorporates trace weights. We then prove a finite approximation theorem that provides bounds to this semantics based on expectations taken over   finite, fixed-length traces. This enables us to frame our semantics within a Feynman-Kac (FK) model, and ensures the consistency of the Particle Filtering (PF) algorithm, an instance of SMC, with respect to our semantics. Building on these results, we introduce VPF, a vectorized version of the PF algorithm tailored to PPGs and our semantics. Experiments conducted with a proof-of-concept implementation of VPF   show  very promising results compared to state-of-the-art PP inference tools.
\end{abstract}
\section{Introduction}\label{sec:intro}
%Probabilistic Programming (PP)  inference, applications.
%
Probabilistic Programming Languages (PPLs) \cite{introGA,introKSS} offer a   systematic approach to define arbitrarily complicated probabilistic models. One is typically interested in   performing \emph{inference} on these models,    given observed data; for example,    finding the posterior distribution of the program's output conditioned on the observed data. Here,  in the context of formal operational semantics of Probabilistic Programs, we study   provably correct and parallelizable  instantiations of  Sequential Monte Carlo (SMC) inference.

%As far as the formal semantics of PPLs is concerned, the denotational approach due to Kozen \cite{Kozen} represents   a well-established mathematical foundation.
%If external observed data are taken into account, a  probabilistic program (PP)   is seen as defining a posterior distribution on variables, given  the observed data.
In terms of  formal semantics of PPLs, the denotational approach introduced by Kozen \cite{Kozen} offers a solid mathematical foundation. However, when it comes to practical algorithms for PPL-based inference, the landscape appears somewhat fragmented. On one hand, \emph{symbolic} and \emph{static analysis} techniques, see e.g. \cite{Sankaranarayanan,psi-solver,Hakaru,BartocciEtAl,OngGuarantees,TribastonePOPL},   yield results with correctness guarantees firmly grounded in the semantics of PPLs but often struggle with scalability.
%When it comes to algorithms for actual PPL-based inference, the situation can be regarded as somewhat unsatisfactory. On  one hand, one finds  \emph{symbolic} or \emph{static analysis} techniques, whose results come equipped with guarantees of correctness,  rigorously  based on the PPL semantics, but   whose  scalability appears to be  quite limited; see e.g. \cite{Sankaranarayanan,psi-solver,Hakaru,BartocciEtAl,OngGuarantees,TribastonePOPL} and references therein.
On the other hand, practical languages and inference algorithms predominantly leverage Monte Carlo (MC) \emph{sampling} techniques (MCMC, SMC), which are more scalable but often lack a clear connection to   formal semantics   \cite{webppl,Stan,Pyro}. Notable exceptions to this situation include works such as \cite{R2,WuEtAl,LundenBorgstrom,VMCAI24,lew}, which are discussed in the related work section.
%On the other hand, there are practical languages and   inference algorithms mostly based  on Monte Carlo (MC) \emph{sampling}  techniques  (MCMC, SMC,...) that have a considerably wider range of applicability, but whose exact relation with  an underlying PPL's formal semantics   is often unclear; see e.g.  \cite{webppl,Stan,Pyro} and references therein (a few notable exceptions to this situation  \cite{R2,WuEtAl,LundenBorgstrom,VMCAI24} are discussed in the Related work section below).

Establishing the consistency of an inference algorithm with respect to a PPL's formal semantics is not merely a theoretical pursuit. In the context of \emph{universal}  PPLs \cite{GoodmanEtal}, integration of unbounded loops and conditioning with MC sampling, which requires truncating computations at a finite time, presents significant challenges \cite{OngGuarantees}. Additionally, the interplay between continuous and discrete distributions in these PPLs can lead to complications, potentially causing existing sampling-based algorithms to yield incorrect results \cite{WuEtAl}.
%Rigorously establishing   consistency of an  inference algorithm  with respect to  a PPL's formal semantics is by no means  of theoretical interest only. Specifically, in the case of \emph{universal} \cite{GoodmanEtal} PPLs, reconciling  unbounded loops  with MC sampling, that requires truncating the  computation at a finite time, can be challenging: see e.g. the examples discussed in \cite{OngGuarantees}. The   mix of continuous and discrete distributions in such PPLs may also give rise to subtleties, and   lead existing sampling-based   algorithms to produce incorrect results: see e.g. the examples discussed in \cite{WuEtAl}.
In the present work, we establish a precise connection between \emph{Probabilistic Program Graphs (PPGs)}, a   general  automata-theoretic    description format of PPs, and \emph{Feynman-Kac (FK)} models, a  formalism for   state-based probabilistic processes and observations defined over a finite time horizon \cite[Ch.5]{SMC}.
This connection enables us to prove the consistency  for PPGs of the \emph{Particle Filtering (PF)} inference algorithm, one of the incarnations of Sequential Monte Carlo approach \cite[Ch.10]{SMC}. In establishing this connection, we adopt a decisively
operational perspective, as explained below.

In a PPG (Section \ref{sec:PP}), computation (essentially, {sampling}) progresses in successive stages specified by the direct edges of a graph (transitions), with nodes serving as {\emph{checkpoints}} between stages for \emph{conditioning} on observed data or more generally updating computation weights.
%
%This will allow  us to prove the consistency for  PPGs of  the \emph{Particle Filtering (PF)}  inference algorithm \cite[Ch.10]{SMC}, one of the incarnations of Sequential Monte Carlo (SMC).
%
%In more detail,   departing  from the traditional denotational approach, we take a  decisively {operational} point of view. In a PPG (Section \ref{sec:PP}), one may think of  computation (basically,    {sampling})  as taking place in successive stages   specified on the direct  edges   of a graph (transitions), with   nodes serving as  {checkpoints} between  stages for \emph{conditioning} on observed data --- or, more generally, for updating the weight assigned to     computations.
%
%Edges also accounts for the control flow among the different stages  via predicates on the store.
The operational semantics of PPGs is formalized in terms of   Markov kernels and score functions.
Building on this, we introduce a measure-theoretic, infinite-trace semantics (Section \ref{sec:obs}, with the necessary measure theory reviewed in Section \ref{sec:measure}). A finite approximation theorem (Section \ref{sec:FA}) then allows us to relate this trace semantics precisely to a finite-time horizon {FK model} (Section \ref{sec:MC}). PF is known to be \emph{consistent} for FK models asymptotically: as the number $N$ of simulated instances (\emph{particles}) tends to infinity, the distribution of these particles converges to the measure defined by the FK model \cite[Ch.11]{SMC}. Therefore,   consistency of PF for PPGs will automatically follow.

%On top of   it, we   introduce a measure theoretic,  infinite-trace semantics (Section \ref{sec:obs}; the necessary measure theory is reviewed in Section \ref{sec:measure}). A finite approximation theorem will then enable us to precisely    relate this trace semantics to a finite-time horizon {FK model} (Section \ref{sec:MC}).   PF has been  shown to be consistent for    FK models,  in an asymptotic sense:  as the number $N$ of simulated instances of the process (\emph{particles})  goes to infinity, the distribution of the  particles   tends to the  measure defined by the FK model \cite[Ch.11]{SMC}. Therefore,  consistency  of PF  for PPGs    will automatically   follow.

Our approach    yields    additional   insights. First,    the finite approximation theorem holds for a class of \emph{prefix-closed functions} defined on infinite traces: these are the functions where the output only depends on a finite initial segment of the input argument.  The finite approximation theorem implies that  the expectation of a  prefix-closed function, defined on the probability space of infinite traces,  can be approximated  by  the expectation  of   functions defined over truncated traces, with respect to a measure  defined  on a suitable    FK model. As expectation in a FK model  can be effectively estimated, via   PF or other algorithms, our finite approximation result lays a sound basis for the statistical model checking of PPs. Second,   the automata-theoretic operational semantics  of PPGs    translates into   a \emph{vectorized}  implementation of PF,  leveraging the fine-grained, SIMD parallelism  existing at the level of particles.
%In particular,  the at each step, the transition function and the combined score function  are applied  at once to the whole vector of $N$ simulated particles.
Specifically, the PPG's transition function and  the score functions are applied simultaneously to the entire vector of $N$ simulated particles at each step.
%\footnote{The \emph{resampling} operation in   PF can only be partially vectorized, though.}.
%This is pragmatically important, as modern CPUs and programming languages offer extensive support for vectorization, and its adoption can lead to dramatic speedups.
This is practically significant, as modern CPUs and programming languages offer extensive support for vectorization, that may lead to dramatic speedups.
We demonstrate this aspect with a prototype vectorized implementation of a PPG-based PF algorithm using TensorFlow \cite{TF}, called VPF. Experiments comparing VPF with state-of-the-art PPLs  on challenging examples from the literature show very promising results (Section \ref{sec:experiments}).
Concluding remarks are provided in the final section (Section \ref{sec:concl}). \iffull Most proofs and additional technical material have been confined to separate appendices (\ref{app:proofs}, \ref{app:PF}, \ref{app:air}).\else Omitted proofs and additional technical material have been reported in an extended version available online \cite{BC25}.\fi
%To concretely demonstrate this aspect, we   consider  a prototype  vectorized implementation of a PPG-based PF algorithm, based on  TensorFlow \cite{TF}, which we call   VPF. Preliminary experiments and comparison with state-of-the-art PPLs   on two challenging examples  show   very encouraging results (Section \ref{sec:experiments}). Some concluding remarks are drawn in the final section (Section \ref{sec:concl}).
%Some technical proofs and additional details on experiments have been confined to the Supplementary Material accompanying this submission   \cite{SM}.

\paragraph{Main contributions}
In summary, our main contributions are as follows.
\begin{enumerate}
\item A clean operational semantics for PPGs, based on expectation taken over infinite-trace,  incorporating conditioning/reweighting.
\item A finite approximation theorem linking this semantics to finite traces and FK models, thereby establishing the consistency of PF for PPGs.
\item A vectorized version of the PF algorithm based on PPGs, and experimental evidence of its practical scalability and competitiveness.
\end{enumerate}

%To sum up, we offer the following   main contributions. (1) A clean, measure-theoretic semantics for   PPGs, a general model of probabilistic programs; (2) a finite approximation theorem that    establishes a link between this semantics and FK models, hence consistency of PF for PPGs; (3) based  on   PPGs, a vectorized version of the PF algorithm.

%\vspace{-0.5cm}
\paragraph{Related work}

With few notable exceptions, most work on the semantics of PPL follows  the denotational approach initiated by Kozen \cite{Kozen}; see \cite{introKSS,GB,borgstrom,introGA,gorinova1,gorinova2,scibior,Staton2,Staton3,joseph}
and references therein for representative works in this area.  In this context, a general goal orthogonal to ours, is devising methods to combine and reason on densities. Note that we
do not require that a PP induces a density on the probability space of infinite traces.

%For instance,  \cite{scibior} provide a denotational validation for advanced inference algorithms for probabilistic programs. They start form \cite{borgstrom}, but work in a modular way with monads, denotational semantics and higher-order functions. Regardless of the adopted method, endowing a language with precise measure-theoretic semantics enables reasoning about the semantics of complicated probabilistic programs: comparing and combining densities defined  with respect to
%different base measures \cite{radul} or performing disintegration on them \cite{narayanan}.

Relevant to our approach is a series of works by Lunden  et al.  on SMC inference applied to PPLs. In \cite{LundenBorgstrom}, for a lambda-calculus enriched with an explicit \textsf{resample} primitive,  consistency of PF is shown to hold, under certain restrictions,   independently of the placements of the \textsf{resample}s in the code. Operationally, their functional approach is very different from our  automata-theoretic one. In particular, they handle suspension and resumption of particles  in correspondence of resampling   via an implicit use of \emph{continuations}, in the style of webPPL \cite{webppl} and other   PPLs. The combination of functional style and continuations does not naturally lend itself to vectorization. For instance, ensuring that all particles are \emph{aligned}, that is are at a  {\textsf{resample}} point of their execution, is an issue that can impact negatively on performance or accuracy.
On the contrary, in  our automata-theoretic model, placement of \textsf{resample}s    and alignment are not issues: resampling always  happens after each  (vectorized) transition step, so all particles are automatically aligned. Note that in PPGs a transition can group together complicated, conditioning free      computations; in any case, consistency of PF is guaranteed. In a subsequent   work \cite{CorePPL,LundenAlign}, Lunden et al. study   concrete implementation issues of SMC. In  \cite{CorePPL}, they  consider \emph{PPL Control-Flow Graphs} (PCFGs), a structure intended as a target for the compilation of high-level PPLs, such as their CorePPL. The PCFG model is very similar in spirit to   PPGs, however, it lacks a formal semantics. Lunden et al. also offer an implementation of this framework,   designed to take advantage of the  potential parallelism existing at the level of particles. We compare our implementation with theirs in Section \ref{sec:experiments}.
%\cite{LundenAlign} --> Automatic Alignment in Higher-Order Probabilistic Programming Languages

%PCFGs are also considered  as the basic operational model   in a series of works by   K. Chatterjee et al., see e.g. \cite{ChatCAV22} and references therein. The operational semantics they consider is similar to ours, but they do not consider an expectation-based semantics, where we incorporate weights. More generally, in  their works  emphasis is on verification, like  obtaining  certified  bounds on  the termination probability   of a given   program. Issues related to conditioning  and to consistency of sampling algorithms are not considered. Wang et al. \cite{Wang24} also consider a model very similar to PCFGs, but their focus   is on exact/symbolic techniques (see further below).

\ifmai
In \cite{hirata}, Hirata et al. propose an Isabelle/HOL library for
probabilistic programs supporting higher-order functions, sampling, and conditioning. However, also in this case the focus is on verification and execution times are not taken into
account,

In \cite{gorinova1}, the authors
provide a syntax and a semantics for Core Stan, a core subset of the Stan language \cite{Stan}, whose semantics has been mainly given in terms of intuition and has not been formalized.
However, this subset omits constraint data types, while loops, random number generators,
recursive user defined functions, and local variables. Morever, they introduce the probabilistic programming language SlicStan.  SlicStan is a more compositional, self-optimizing version
of Stan, but also in this case it does not support while loops and recursive functions, making a small restriction compared to Stan.
SlicStan has been subsequently extended in \cite{gorinova2} to support discrete parameters in the case when the discrete parameter space is bounded.
\fi

%Many probabilistic programming languages and libraries \cite{Stan,cusumano,gorinova1,gorinova2,hirata,joseph} have been defined recently, % such as Stan, SlicStan, Gen or Isabelle/HOL library,
% in some cases equipped with a formalization of the intended semantics.
%SMCP3 \cite{lew} gives inference algorithm designers a more flexible framework than Gen \cite{cusumano} and provides a precise measure-theoretic formalism for an extension of the Gen language. However, often the focus is more on verification or on the choice of expressive proposals, instead we aim to show a connection between the language and its formal semantics.
Aditya et al. prove consistency of Markov Chain Monte Carlo (MCMC) for their PPL R2
\cite{R2}, which is based on a big-step    sampling
semantics that considers finite execution paths. No
approximation results bridging  finite and infinite traces, and hence unbounded loops, is
provided. It is also  unclear if  a big-step semantics would effectively
translate  into a SIMD-parallel algorithm. Wu et al. \cite{WuEtAl} provide the PPL Blog with a rigorous measure-theoretic semantics, formulated   in terms of Bayesian Networks, and a very efficient implementation of the PF algorithm   tailored to such networks. Again, they do not offer results for unbounded loops. In our previous work \cite{VMCAI24},
we have considered  a measure theoretic semantics for a PPL with unbounded loops, and provided a finite approximation result   and a SIMD-parallel implementation, with guarantees,  of what is in effect a \emph{rejection sampling} algorithm. Rejection may be effective for limited forms of conditioning; but it rapidly becomes wasteful and ineffective as   conditioning becomes more demanding, so to speak: e.g. when it is repeated in a loop, or the observed data have a low likelihood in the model. Finally, SMCP3 \cite{lew} provides a  rich   measure-theoretic framework for extending  the practical Gen language \cite{cusumano} with expressive proposal distributions.

A rich area in the field of PPL focuses on symbolic, exact techniques \cite{Sankaranarayanan,psi-solver,Hakaru,BartocciEtAl,OngGuarantees,TribastonePOPL,hirata} aiming to obtain  termination certificates, or certified bounds on termination probability of PPs, or even exact representations of the posterior distribution; see also  \cite{1,2,3,4,5,Wang24} for some recent works in this direction. Our goal and methodology, as already stressed, are rather different, as we focus on scalable inference
via sampling
%with statistical guarantees,
and the ensuing consistency issues. This is part of a broader research agenda, aimed at developing flexible and scalable formal methods applicable  across   diverse, probability-related domains,  including: dynamical  systems with safety-related aspects \cite{HSCC18,SOFSEM18,LMCS19,MFCS19,IC22,VMCAI23}, information leakage and security \cite{ISC14,QIFForte14}, distributed  systems with notions of failure and recovery \cite{CaspisMSCS},   randomized model counting and  testing \cite{BG19,OutgenIEEE}.

%Also, no
%approximation results in terms of finite execution paths is
%provided.

%In this line of work emphasis is, for instance,  on providing conditions under which a
%\emph{density} for a program-induced random variable exists. We
%do not consider such aspects in our framework, as a
%(cumulative) \emph{distribution}, which always exists, is all
%that is needed.

\section{Preliminaries on measure theory}\label{sec:measure}
We review a few basic concepts from measure theory following closely the presentation in the first two chapters of  \cite{Ash}, which is a reference for whatever is not explicitly described below.
\begin{comment}
	, recall the concepts of: sigma-field,  Borel sets, Borel sigma-field, measurable space, (sigma-finite)  measure,  Borel measurable function, Lebesgue  measures, Dirac's measure,  discrete measure, counting measure.
	In what follows,   measurable will always mean \textbf{\emph{Borel measurable}}.
	Recall basic facts about Borel measurable functions, in particular: Borel measurability is preserved by composition and other elementary operations on functions; continuous real functions are Borel measurable.
\end{comment}
Given a nonempty set $\Omega$, a \emph{sigma-field} $\F$ on $\Omega$ is a collection of subsets of $\Omega$ that contains $\Omega$, and is closed under complement and under countable disjoint union. The pair $(\Omega,\F)$ is called a \emph{measurable space}. A (total) function $f:\Omega_1\rightarrow \Omega_2$ is \emph{measurable} w.r.t. the sigma-fields $(\Omega_1,\F_1)$ and $(\Omega_2,\F_2)$ if whenever $A\in \F_2$ then $f^{-1}(A)\in \F_1$.
We let $\overline\reals=\reals\cup\{-\infty,+\infty\}$ be the set of extended reals, assuming the standard arithmetic for $\pm\infty$ (cf. \cite[Sect.1.5.2]{Ash}), and $\ereals^+$ the set of nonnegative reals including $+\infty$. The \emph{Borel sigma-field} $\F$ on $\Omega=\ereals^m$ is the minimal sigma-field that contains all {rectangles} of the form $[a_1,b_1]\times\cdots \times [a_n,b_n]$, with $a_i,b_i\in \ereals$.  An important case of measurable spaces $(\Omega,\F)$ is when $\Omega=\ereals^m$  for some $m\geq 1$  and  $\F$ is the Borel sigma-field over $\Omega$. Throughout the paper, {\emph{“measurable” means “Borel measurable”}}, both for sets and for functions. On functions, Borel measurability is preserved by composition and other elementary operations on functions; continuous real functions are Borel measurable. We will let $\F_k$ denote   the Borel sigma-field over $\overline\reals^{k}$ ($k\geq 1$) when we want to be specific about the dimension of the space.

A \emph{measure} over a measurable space $(\Omega,\F)$ is a function $\mu:\F\rightarrow \ereals^+$ that is countably additive, that is $\mu(\cup_{j\geq 1}A_j)=\sum_{j\geq 1}\mu(A_j)$ whenever $A_j$'s are pairwise disjoint sets in $\F$.
%
%We shall only work with measurable spaces $(\Omega,\F)$ where $\Omega=\ereals^m$  for some $m\geq 1$  and  $\F$ is the Borel sigma-field on $ \ereals^m$. In particular, recall that the Borel sigma-field on $\ereals^m$ is the minimal sigma-field that contains $\R$, the set of \emph{rectangles} of the form $[a_1,b_1]\times\cdots \times [a_n,b_n]$, with $a_i,b_i\in \ereals$.
The \emph{Lebesgue integral} of a Borel measurable function $f$  w.r.t. a measure $\mu$ \cite[Ch.1.5]{Ash}, both  defined over a measure space  $(\Omega,\F)$, is denoted by $\int_\Omega \mu(d\omega)f(\omega)$,
with the subscript $\Omega$ omitted when clear from the context. When $\mu$ is the standard Lebesgue measure, we  may omit $\mu$ and write the integral   as   $\int_\Omega d\omega f(\omega)$.  For $A\in \F$, $\int_A \mu(d\omega)f(\omega)$ denotes $\int_\Omega \mu(d\omega)f(\omega)1_A(\omega)$, where $1_A(\cdot)$ is the indicator function of the set $A$. We let $\delta_v$ denote  Dirac's measure concentrated on $v$: for each  set $A$ in an appropriate sigma-field,  $\delta_v(A)=1$ if $v\in A$,   $\delta_v(A)=0$ otherwise. Otherwise said, $\delta_v(A)=1_A(v)$. Another measure that arises (in connection with discrete distributions) is the counting measure,  $\mu_C(A):=|A|$. In particular, for a nonnegative $f$, we have the equality $\int_A \mu_C(d\omega)f(\omega)=\sum_{\omega\in A}f(\omega)$.
%In particular, for $v=(v_1,...,v_m)\in\overline\reals^m$, $\delta_v$ can be expressed as a product measure: $\delta_v=\delta_{v_1}\times\cdots \times \delta_{v_m}$.
%
%
%;  product space, Fubini theorem...(what else?)
%
A \emph{probability measure} is a   measure $\mu$ defined on $\F$ such that $\int \mu(du)=1$.
%Note that probability measures are trivially  finite measures, hence sigma-finite a fortiori.
For a given nonnegative measurable function $f$ defined over $\Omega$, its \emph{expectation} w.r.t. a probability measure $\nu$  is just its integral: $\expc_\nu[f]=\int\nu(d \omega)f(\omega)$.
%Over $\reals^m$, we shall often make  use of the Lebesgue product measure $\mu^m:=\mu\times\cdots \times \mu$ ($m$ times), where $\mu$ is the standard Lebesgue measure over $\reals$. Both  $\mu^m$  and $\delta_y$ are   Lebesgue-Stiltijes measures (over the appropriate measurable spaces).
The following definition is central.
\begin{definition}[Markov kernel]\label{def:MK} Let $(\Omega_1,\F_1)$ and $(\Omega_2,\F_2)$ be   measurable spaces. A function $K: \Omega_1\times \F_2 \longrightarrow \erealspl$ is a \emph{Markov kernel}  from $\Omega_1$ to $\Omega_2$  if  it satisfies the following properties:
	\begin{enumerate}
		\item for each $\omega\in \Omega_1$, the function $K(\omega,\cdot):\F_2\rightarrow \erealspl$ is a probability measure on $(\Omega_2,\F_2)$;
		\item for each $A\in \F_2$, the function $K(\cdot,A): \Omega_1 \rightarrow \erealspl$ is   measurable.
	\end{enumerate}
\end{definition}

Notationally, we will most often write $K(\omega,A)$ as $K(\omega)(A)$.
%We will mostly be concerned with the case $\Omega_1=\Omega_2$, $\F_1=\F_2$.
The following is a standard result about  the  construction  of finite product of measures  over a  product space\footnote{We shall freely identify language-theoretic \emph{words} with \emph{tuples}.}
%, hence use the notations  $A_1\cdot A_2 \cdot\,\cdots\, \cdot A_k$ and  $A_1\times A_2\times \cdots \times A_k$ interchangeably. This convention will also apply to infinite words (cf. Section \ref{sec:obs}).
  $\Omega^t=\Omega\times \cdots \times \Omega$ ($t$ times) for $t\geq 1$ an integer\iffull (see Theorem \ref{th:prode} in the Appendix for a more detailed statement).\else. \fi It is customary to denote the measure $\mu^t$ defined by  the theorem also as $\mu^1\otimes K_2\otimes\cdots \otimes K_t$.

\begin{theorem}[product of measures, \cite{Ash}, Th.2.6.7]\label{th:prod}
	Let  $t\geq 1$ be an integer. Let $\mu^1$ be a probability measure on $\Omega$ and $K_2,...,K_t$ be $t-1$   (not necessarily distinct) Markov kernels from $\Omega$ to $\Omega$. Then there is a unique probability measure $\mu^t$ defined on $(\Omega^t,\F^t)$ such that for every $A_1\times \cdots\times A_t\in \F^t$ we have:
	$\mu^t(A_1\times \cdots\times A_t) = \int_{A_1} \mu^1(d\omega_1)\int_{A_2}K_2(\omega_1)(d\omega_2)\cdots \int_{A_t}K_t(\omega_{t-1})(d\omega_t)
	$.
\end{theorem}

\section{Probabilistic programs}\label{sec:PP}
We first  introduce a general formalism for specifying programs, in the form of certain graphs  that can be regarded as symbolic finite automata. For this formalism, we introduce then an operational semantics  in terms of  Markov kernels. %Finally, we also outline a minimalistic but hopefully    handy language to specify programs.
%\vspace{-0.5cm}
\paragraph{Probabilistic Program Graphs}\label{sec:PPG}
%First,  writing programs, one
In defining probabilistic programs, we will   rely on  a repertoire of basic distributions: % We introduce the necessary terminology below.
%
%We will assume that each   basic distribution  admits a  density w.r.t. a suitable measure;
continuous, discrete and mixed distributions will be allowed. A crucial point for   expressiveness  is that a measure may depend on  \emph{parameters}, whose   value at runtime is determined by the state of the program. To ensure that the resulting programs  define measurable functions (on a suitable space), it is important that the dependence between the parameters and the measure be in turn of measurable type.  We will formalize this in terms of Markov kernels. Additionally, we will   consider score functions, a generalization of 0/1-valued predicates.
%
%Before introducing the actual syntax of our language, We introduce the syntax of probabilistic programs.
Formally, we will consider the     two families of functions defined below. In the definitions, we will let  $m\geq 1$ denote a fixed integer, representing the number of \emph{variables} in the program,   conventionally referred  to as $x_1,...,x_m$. We will let $v$ range over $\ereals^m$,  the content of the program variables in a given state, or \emph{store}.
\begin{itemize}
	%\item \emph{Update functions}:     measurable    functions  $g:\ereals^m\rightarrow \ereals$.
	%\mik{Qui, controllare uso counting measure.}
	\item \emph{Parametric measures}:   Markov kernels       $\zeta:\ereals^m \times \F_m \rightarrow [0,1] $.
	%For each such $G$, we will assume a measure $\mu_G$ over $\ereals$ is given, such that the function $(v,A)\mapsto \int_A \rho(r,v)\mu_G(d r)  $ is a Markov kernel from $\ereals^m$ to $\ereals$.
	%   such that for each $v \in \reals^m$, $G(\cdot, v)$ is a density w.r.t. a measure $\mu_G$ defined $\ereals$ --- that is, for each $v$,   $\int_{\ereals}  G(r, v)\mu_G(dr) =1$.
	\item \emph{Score functions}:    measurable functions  $\gamma:\ereals^m\rightarrow [0,1]$.
	%By $\overline\gamma$ we denote the score function $1-\gamma$.
	A \emph{predicate} is a special case of a score function $\varphi:\ereals^m\rightarrow \{0,1\}$. An Iverson bracket style notation will be often employed, e.g.: $[x_1\geq 1]$ is the   predicate  that on input $v$ yields 1 if $v_1\geq 1$, 0 otherwise.  %Indicator functions for measureable sets $A$, denoted $1_A$, are
\end{itemize}
For a parametric measure $\zeta$ and a store $v\in \ereals^m$,  $\zeta(v)$  is a distribution, that can be used to  sample a new store  $v'\in \ereals^m$   depending on the current program store $v$. Analytically, $\zeta$   may be expressed by,  for instance,   chaining together sampling of individual components of the store. This can be done by relying on \emph{parametric densities}:  measurable functions $\rho:\ereals^m\times \ereals \rightarrow \ereals^+$ such that, for a designated measure $\mu_\rho$, the function
$(v,A)\mapsto \int_A\mu_\rho(dr)\,\rho(v,r)$ ($A\in \F_m$) is a Markov kernel from $\ereals^m$ to $\ereals$. This is explained via the following example.

\begin{example}{\label{ex:parametricD}
		Fix $m=2$. Consider the Markov kernel defined as follows, for each $x_1,x_2\in \ereals$ and $A\in \F_2$
		\vspace{-0.2cm}
		\begin{equation}\label{eq:zeta}
			\begin{array}{c}
				\zeta(x_1,x_2)(A):=\int \mu_1(d r_1)\, \left(\rho_1(x_1,x_2,r_1)\cdot \int\mu_2(d r_2)\rho_2(r_1,x_2,r_2)1_A(r_1,r_2)\right)
			\end{array}
		\end{equation}
		\noindent
		where:  $\mu_1=\mu_C$ is the counting measure;  $\rho_1(x_1,x_2,r)=\frac 1 2 1_{\{x_1\}}(r)+\frac 1 2 1_{\{x_2\}}(r)$ is the density of a discrete distribution on $\{x_1,x_2\}$;  $\mu_2=\mu_L$ is the ordinary Lebesgue measure; $\rho_2(x_1,x_2,r)=N(x_1, x_2 ,r):=\frac 1 { |x_2|\sqrt{2\pi} }\exp(-\frac 1 2 (\frac{ r-x_1}{|x_2|} )^2 )$ is the  density of the Normal distribution of mean $x_1$ and standard deviation\footnote{With the proviso that,  when $x_2=0$ or $|x_1|,|x_2|=+\infty$, $N(x_1,x_2,r)$ denotes an arbitrarily fixed, default probability density.} $|x_2|$.
		The function $\zeta$ is a parametric measure: concretely, it corresponds to  first sampling uniformly $r_1$ from the set $\{x_1,x_2\}$,   then sampling $r_2$ from the Normal distribution of mean $r_1$ and s.d. $|x_2|$ (if $|x_2|$ is positive and finite, otherwise from a default distribution). Rather than via \eqref{eq:zeta}, we will describe $\zeta$   via the following more handy notation: $r_1\sim \rho_1(x_1,x_2)\,;\,r_2\sim \rho_2(r_1,x_2)$
%		$$
%		\begin{array}{c}
%			r_1\sim \rho_1(x_1,x_2)\,;\,r_2\sim \rho_2(r_1,x_2)
%			%\zeta(x_1,x_2)(A):=\int \mu_1(d r_1)\, \left(\rho_1(x_1,x_2,r_1)\cdot \int\mu_2(d r_2)\rho_2(r_1,x_2,r_2)1_A(r_1,r_2)\right)
%		\end{array}
%		$$
		(or listed top-down). Note that the sampling order from left to right is relevant here.
}\end{example}

In fact, as far as the formal framework of PPGs introduced below is concerned, how the parametric measures $\zeta$'s are analytically described is irrelevant. From the practical point of view, it is important   we know how to (efficiently) sample from the measure $\zeta(v)$,  for any $v$, in order for  the inference algorithms to be actually implemented (see Section \ref{sec:MC}). In concrete terms,    $\zeta(v)$ might represent  the (possibly unknown) distribution of the outputs in $\ereals^m$ returned by a piece of code, when invoked with input $v$.
%Moreover, we let $v_{\text{-}i}\in \ereals^{m-1}$ be the vector obtained from $v$ by removing its $i$-th component, and let $A_{(v_{\text{-}i},S)}=\{r\in \ereals: (v[r@i],S)\in A\}\subseteq \ereals$  be the section of $A$  at  $(v_{\text{-}i},S)$, which is a measurable set in $\F_1$.
Another important special case  of parametric measure is the following. For any   $v=(v_1,...,v_m)\in \overline\reals^m$,  $r\in \ereals$ and $1\leq i \leq m$,   let $v[r@ i]:=(v_1,...,r,...,v_m)$ denote  the tuple where $v_i$ has been replaced by $r$.  Consider the parametric measure  $\zeta(v) = \delta_{v[g(v)@i]}$, where $g:\ereals^m\rightarrow \ereals$ is a measurable function. In programming terms, this corresponds to the deterministic  \emph{assignment} of the value $g(v)$ to the variable $x_i$. We will describe this $\zeta$ as: $x_i:=g(x_1,...,x_m)$.

In the definition of PPG below, one may think of   the computation (sampling) taking place in successive stages    on the  edges (transitions) of the graph, with   nodes serving as \emph{checkpoints} (a term we have borrowed from \cite{LundenBorgstrom})  between  stages for conditioning on observed data --- or, more generally, re-weighting the score assigned to a computation. The edges also account for the control flow among the different stages  via predicates computed on the store of the source nodes.

\lstset{
basicstyle=\ttfamily,
mathescape
}

\begin{figure}[t!]
\begin{minipage}{.17\linewidth}
%\begin{center}
\vspace{-0.25cm}
{\scriptsize
	\begin{lstlisting}
c$\sim$B(1/2);
if (c==0) skip
else {
    d$\sim$B(1/2);
    observe(d==1);
    skip
}
	\end{lstlisting}
}
\vspace{-0.3cm}
%\end{center}	
\end{minipage}
%\hspace*{.1cm}
\begin{minipage}{.22\linewidth}
{\small
	\begin{center}
		\tikzset{every picture/.style={line width=0.75pt}} %set default line width to 0.75pt	
		\begin{tikzpicture}[x=0.75pt,y=0.75pt,yscale=-.7,xscale=.7]
			%uncomment if require: \path (0,300); %set diagram left start at 0, and has height of 300
			%			
			%Shape: Circle [id:dp2626517041288785]
			\draw   (185.13,65.54) .. controls (185.13,72.4) and (190.69,77.96) .. (197.54,77.96) .. controls (204.4,77.96) and (209.96,72.4) .. (209.96,65.54) .. controls (209.96,58.68) and (204.4,53.12) .. (197.54,53.12) .. controls (190.69,53.12) and (185.13,58.68) .. (185.13,65.54) -- cycle ;
			%Shape: Circle [id:dp15208927176210318]
			\draw   (185.13,125.54) .. controls (185.13,132.4) and (190.69,137.96) .. (197.54,137.96) .. controls (204.4,137.96) and (209.96,132.4) .. (209.96,125.54) .. controls (209.96,118.68) and (204.4,113.12) .. (197.54,113.12) .. controls (190.69,113.12) and (185.13,118.68) .. (185.13,125.54) -- cycle ;
			%Shape: Circle [id:dp6511223174619254]
			\draw   (233.13,176.54) .. controls (233.13,183.4) and (238.69,188.96) .. (245.54,188.96) .. controls (252.4,188.96) and (257.96,183.4) .. (257.96,176.54) .. controls (257.96,169.68) and (252.4,164.12) .. (245.54,164.12) .. controls (238.69,164.12) and (233.13,169.68) .. (233.13,176.54) -- cycle ;
			%Shape: Circle [id:dp5133119840513822]
			\draw   (140.56,176.85) .. controls (140.56,183.71) and (146.12,189.27) .. (152.98,189.27) .. controls (159.84,189.27) and (165.4,183.71) .. (165.4,176.85) .. controls (165.4,170) and (159.84,164.44) .. (152.98,164.44) .. controls (146.12,164.44) and (140.56,170) .. (140.56,176.85) -- cycle ;
			%Shape: Circle [id:dp007214209928751103]
			\draw   (143.25,176.85) .. controls (143.25,182.23) and (147.61,186.58) .. (152.98,186.58) .. controls (158.35,186.58) and (162.71,182.23) .. (162.71,176.85) .. controls (162.71,171.48) and (158.35,167.12) .. (152.98,167.12) .. controls (147.61,167.12) and (143.25,171.48) .. (143.25,176.85) -- cycle ;
			%Straight Lines [id:da06486395183775184]
			\draw    (196.69,39.23) -- (197.36,50.13) ;
			\draw [shift={(197.54,53.12)}, rotate = 266.49] [fill={rgb, 255:red, 0; green, 0; blue, 0 }  ][line width=0.08]  [draw opacity=0] (8.93,-4.29) -- (0,0) -- (8.93,4.29) -- cycle    ;
			%Straight Lines [id:da3535869522546664]
			\draw    (197.54,77.96) -- (197.54,110.12) ;
			\draw [shift={(197.54,113.12)}, rotate = 270] [fill={rgb, 255:red, 0; green, 0; blue, 0 }  ][line width=0.08]  [draw opacity=0] (8.93,-4.29) -- (0,0) -- (8.93,4.29) -- cycle    ;
			%Straight Lines [id:da6766977121884594]
			\draw    (206.5,135.6) -- (235.45,166.41) ;
			\draw [shift={(237.5,168.6)}, rotate = 226.79] [fill={rgb, 255:red, 0; green, 0; blue, 0 }  ][line width=0.08]  [draw opacity=0] (8.93,-4.29) -- (0,0) -- (8.93,4.29) -- cycle    ;
			%Straight Lines [id:da7468869553086717]
			\draw    (188.5,135.6) -- (163.42,166.79) ;
			\draw [shift={(161.54,169.12)}, rotate = 308.8] [fill={rgb, 255:red, 0; green, 0; blue, 0 }  ][line width=0.08]  [draw opacity=0] (8.93,-4.29) -- (0,0) -- (8.93,4.29) -- cycle    ;
			%Straight Lines [id:da564880834563144]
			\draw    (233.13,176.54) -- (168.4,176.84) ;
			\draw [shift={(165.4,176.85)}, rotate = 359.74] [fill={rgb, 255:red, 0; green, 0; blue, 0 }  ][line width=0.08]  [draw opacity=0] (8.93,-4.29) -- (0,0) -- (8.93,4.29) -- cycle    ;
			%Curve Lines [id:da2395648249982678]
			\draw    (147.5,163.6) .. controls (133.02,137.54) and (112.03,178.54) .. (137.59,177.17) ;
			\draw [shift={(140.56,176.85)}, rotate = 171.03] [fill={rgb, 255:red, 0; green, 0; blue, 0 }  ][line width=0.08]  [draw opacity=0] (8.93,-4.29) -- (0,0) -- (8.93,4.29) -- cycle    ;
			%Straight Lines [id:da09289555832352314]
			%\draw  [dash pattern={on 4.5pt off 4.5pt}]  (257,184) -- (283.5,204.8) ;
			
			% Text Node
			\draw (191,58) node [anchor=north west][inner sep=0.75pt]    {$0$};
			% Text Node
			\draw (192,119) node [anchor=north west][inner sep=0.75pt]    {$1$};
			% Text Node
			\draw (240,170) node [anchor=north west][inner sep=0.75pt]    {$3$};
			% Text Node
			\draw (147,170) node [anchor=north west][inner sep=0.75pt]{$2$};
			% Text Node
			\draw (210,77) node [anchor=north west][inner sep=0.75pt]  [font=\scriptsize]  {${\textstyle c\sim B\left(\frac{1}{2}\right)}$};
			% Text Node
			\draw (225,123) node [anchor=north west][inner sep=0.75pt]  [font=\scriptsize]  {$ \begin{array}{l}
					[{\textstyle c\neq 0}],\\
					{\textstyle d\sim B\left(\frac{1}{2}\right)}
				\end{array}$};
			% Text Node
			\draw (140,128) node [anchor=north west][inner sep=0.75pt]  [font=\footnotesize]  {$[{\scriptstyle c=0}]$};
			% Text Node
			\draw (258,178) node [anchor=north west][inner sep=0.75pt]  [font=\scriptsize]  {$\gamma =[d=1]$};

		\end{tikzpicture}	
\end{center}}
\end{minipage}\hspace*{.5cm}
\begin{minipage}{.25\linewidth}
%\begin{center}
\vspace{-0.3cm}
{\scriptsize
	\begin{lstlisting}
d$\sim$U(0,2); r$\sim$U(0,1);
x:=-1; y:=1;
while(|x-y|$\geq$.1) {
     x$\sim$N(x,d); y$\sim$N(y,r);
     observe(|x-y|$\leq$3);
}
	\end{lstlisting}
}
\vspace{-0.3cm}
%\end{center}	
\end{minipage}
\begin{minipage}{.2\linewidth}
{\small
	\begin{center}
		\tikzset{every picture/.style={line width=0.75pt}} %set default line width to 0.75pt
		\begin{tikzpicture}[x=0.75pt,y=0.75pt,yscale=-.7,xscale=.7]
			%uncomment if require: \path (0,265); %set diagram left start at 0, and has height of 265
			%Shape: Ellipse [id:dp8654542714682785]
			\draw   (221.39,117.28) .. controls (221.49,110.04) and (227.23,104.26) .. (234.2,104.37) .. controls (241.17,104.48) and (246.73,110.44) .. (246.62,117.69) .. controls (246.52,124.93) and (240.78,130.71) .. (233.81,130.6) .. controls (226.84,130.49) and (221.28,124.53) .. (221.39,117.28) -- cycle ;
			%Straight Lines [id:da150259421804795]
			\draw    (246.62,117.69) -- (301.63,118.26) -- (327.5,118.26) ;
			\draw [shift={(330.5,118.26)}, rotate = 180] [fill={rgb, 255:red, 0; green, 0; blue, 0 }  ][line width=0.08]  [draw opacity=0] (8.93,-4.29) -- (0,0) -- (8.93,4.29) -- cycle    ;
			%Shape: Ellipse [id:dp33541890147797315]
			\draw   (330.5,118.26) .. controls (330.5,110.25) and (336.74,103.76) .. (344.45,103.76) .. controls (352.15,103.76) and (358.4,110.25) .. (358.4,118.26) .. controls (358.4,126.26) and (352.15,132.76) .. (344.45,132.76) .. controls (336.74,132.76) and (330.5,126.26) .. (330.5,118.26) -- cycle ;
			%Curve Lines [id:da7876754673421058]
			\draw    (358.4,118.26) .. controls (396.51,104.16) and (341.53,67.77) .. (344.13,101.06) ;
			\draw [shift={(344.45,103.76)}, rotate = 261.3] [fill={rgb, 255:red, 0; green, 0; blue, 0 }  ][line width=0.08]  [draw opacity=0] (8.93,-4.29) -- (0,0) -- (8.93,4.29) -- cycle    ;
			%Straight Lines [id:da9148680522623549]
			\draw    (344.45,132.76) -- (344.91,180.87) ;
			\draw [shift={(344.93,182.87)}, rotate = 269.45] [color={rgb, 255:red, 0; green, 0; blue, 0 }  ][line width=0.75]    (10.93,-3.29) .. controls (6.95,-1.4) and (3.31,-0.3) .. (0,0) .. controls (3.31,0.3) and (6.95,1.4) .. (10.93,3.29)   ;
			%Shape: Ellipse [id:dp0910889155936161]
			\draw   (330.98,197.37) .. controls (330.98,189.36) and (337.22,182.87) .. (344.93,182.87) .. controls (352.63,182.87) and (358.88,189.36) .. (358.88,197.37) .. controls (358.88,205.37) and (352.63,211.87) .. (344.93,211.87) .. controls (337.22,211.87) and (330.98,205.37) .. (330.98,197.37) -- cycle ;
			%Shape: Ellipse [id:dp4276976064297249]
			\draw   (333.18,197.37) .. controls (333.18,190.92) and (338.44,185.7) .. (344.93,185.7) .. controls (351.42,185.7) and (356.68,190.92) .. (356.68,197.37) .. controls (356.68,203.81) and (351.42,209.04) .. (344.93,209.04) .. controls (338.44,209.04) and (333.18,203.81) .. (333.18,197.37) -- cycle ;
			%Curve Lines [id:da23449245382564343]
			\draw    (358.88,197.37) .. controls (390.13,196.49) and (368.8,160.14) .. (355.82,185.51) ;
			\draw [shift={(354.64,188.04)}, rotate = 292.96] [fill={rgb, 255:red, 0; green, 0; blue, 0 }  ][line width=0.08]  [draw opacity=0] (8.93,-4.29) -- (0,0) -- (8.93,4.29) -- cycle    ;
			
			% Text Node
			\draw (259.34,43.82) node [anchor=north west][inner sep=0.75pt]  [font=\scriptsize,rotate=-359.58,xslant=0]  {$ \begin{array}{l}
					d\sim U( 0,2) ;\\
					r\sim U( 0,1) ;\\
					x:=-1;\\
					y:=1
				\end{array}$};
			% Text Node
			\draw (375.69,60.58) node [anchor=north west][inner sep=0.75pt]  [font=\scriptsize,rotate=-359.84,xslant=0]  {$ \begin{array}{l}
					[|x-y|\geq 0.1] ,\\
					x\sim N( x,d) ;\\
					y\sim N( y,r)
				\end{array}$};
			% Text Node
			\draw (245.62,150.52) node [anchor=north west][inner sep=0.75pt]  [font=\scriptsize,rotate=-359.84,xslant=0]  {$[|x-y|< 0.1]$};
			% Text Node
			\draw (229.23,109.42) node [anchor=north west][inner sep=0.75pt]    {$0$};
			% Text Node
			\draw (339.55,110.29) node [anchor=north west][inner sep=0.75pt]    {$1$};
			% Text Node
			\draw (339.55,189.62) node [anchor=north west][inner sep=0.75pt]    {$2$};
			% Text Node
			\draw (374.76,174.6) node [anchor=north west][inner sep=0.75pt]  [font=\footnotesize]  {};%$\mathrm{id}$
			% Text Node
			\draw (354.99,122.04) node [anchor=north west][inner sep=0.75pt]  [font=\scriptsize]  {$\gamma =[ \ |x-y|\leq 3\ ]$};
		\end{tikzpicture}		
\end{center}}
\end{minipage}
\vspace{-0.3cm}
\caption{\scriptsize \textbf{Left}. The PPG of Example \ref{ex:simple0} and a corresponding pseudo-code. The    $\nil$ node (2) is distinguished with a double border. Constant 1 predicates and score functions, and the identity function are not displayed in transitions. The score function $\gamma$ decorates node 3, that is $\psco(3)=\gamma$. \textbf{Right}. The PPG for the drunk man and   mouse random walk of Example \ref{ex:dm1} and a corresponding pseudo-code.   The score function $\gamma$ decorates node 1, that is $\psco(1)=\gamma$.}\label{fig:drunk}
\vspace{-0.5cm}
\end{figure}
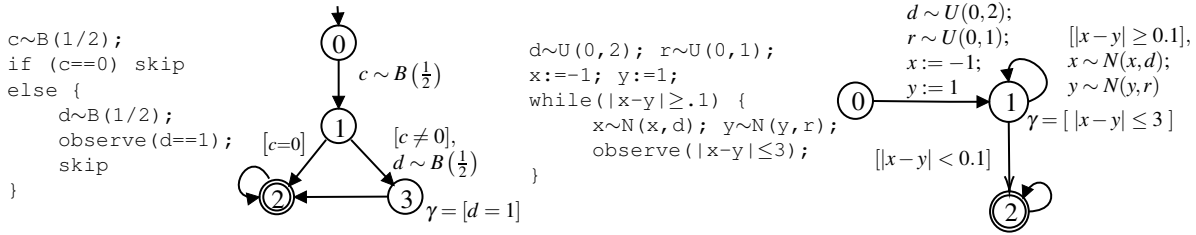

%---------------------------------------------------------------------

\begin{definition}[PPG]\label{def:PPG}
Fix $m\geq 1$. A \emph{Probabilistic Program Graph (PPG)} on $\ereals^m$ is a 4-tuple $\ppg=(\St,E,\nil,\psco)$ satisfying the following.
\begin{itemize}
\item $\St=\{S_1,...,S_k\}$ is a finite, nonempty set of \emph{program checkpoints} (\emph{programs}, for short).
\item $E$ is a finite, nonempty set of \emph{transitions} of the form $(S,\varphi,\zeta,S')$, where: $S,S'\in \St$ are called the \emph{source} and \emph{target} program checkpoint, respectively; $\varphi:\ereals^m\rightarrow\{0,1\}$ is a predicate;  and  $\zeta:\ereals^m \times \F_m \rightarrow [0,1] $ is a parametric measure.
\item $\nil\in \St$ is a distinguished \emph{terminated} program checkpoint, such that $(\nil,1,\mathrm{id},\nil)$ ($\mathrm{id}$ = identity) is the only transition in $E$ with $\nil$ as source.
\item $\psco$ is a mapping from $\St$ to the set of score functions, s.t. $\psco(\nil)$ is the constant  1.
\end{itemize}
Additionally,   denoting by $E_S$ the set of transitions in $E$ with $S$ as a source checkpoint, the following \emph{consistency} condition is   assumed: for each $S\in \St$, the function $\sum_{(S,\varphi,\zeta,S')\in E_S}\varphi $  is  the constant 1.
\end{definition}

We first illustrate Definition \ref{def:PPG} with a simple example. This will also serve to illustrate the finite approximation theorem in the next section.

\begin{example}\label{ex:simple0}
Consider the PPG in Fig. \ref{fig:drunk}, left. Here we have $m=2$    and $B(p)$ is the Bernoulli distribution with success probability $p$.  On the left, a more conventional pseudo-code      notation for the resulting program.
% might be described as follows, where \verb"skip" represents termination.
%\begin{center}
%			\vspace{-0.3cm}
%{\small
%	\begin{lstlisting}
	%		c$\sim$B(1/2);
	%		if (c==0) skip
	%		else {d$\sim$B(1/2); observe(d==1); skip}
	%	\end{lstlisting}
%}
%		\vspace{-0.3cm}
%\end{center}
We will not pursue a systematic formal translation from this program notation to PPGs, though.
\end{example}

%Note that for each program point $S$, the set of transitions $E_S$ is partitioned as $E_S=E^u_{S}\cup E^d_{S}$, where $E^u_{S}$ (resp. $E^d_{S}$) denotes the subset of transitions where $f=g$ is an update function (resp. $f=G$ is a parametric density). These two subsets correspond to \emph{assignments} and sample \emph{drawings} transitions from program point $S$, respectively.
%%%%%%%%%%%%%%%%%%%%%%%%%%%%%%
The following example illustrates the use of scoring functions inside loops. While a bit contrived, it is close to the structure of more significant scenarios, such as the aircraft tracking example of \cite{WuEtAl}, cf. Section \ref{sec:experiments}.

\begin{example}[Of mice and drunk men]{\label{ex:dm1}
On  a street, a  drunk man and a mouse perform independent  random walks  starting at conventional positions $-1$ and $1$ respectively. Initially, each of them samples   a standard deviation (s.d.) from a uniform distribution.
 Then, at each discrete time step, they independently sample  their own next position from a Normal distribution centered at the current position with  the  s.d. chosen at the beginning. The process is stopped as soon as the man and the mouse meet, which we take to mean the distance between them is $<1/10$.

%https://www.theguardian.com/cities/2014/feb/13/urban-myths-6ft-from-a-rat
It has been suggested that  in certain urban areas   a man is never more than 3m away from a   mouse \cite{TheGuardian}. Taking this    information at face value, we incorporate it into   our model  with the score function $\gamma:=[|x-y|\leq 3]$. The resulting PPG is described in Figure \ref{fig:drunk}, right (cf. also pseudo-code).
%Consider the PPG in Figure \ref{fig:drunk}.
}\end{example}

\iffull
\begin{remark}{ %(a)
The definition of score function  requires that $\gamma$ takes values on $[0,1]$. This is restriction is necessary  for the weight function to be defined on infinite traces (cf. \eqref{eq:wt}) be well-defined. A similar restriction is found in e.g. \cite{borgstrom}. In practice, provided that $\gamma$ is nonnegative and bounded, we can always divide by an appropriate constant\footnote{Provided all the score functions appearing in the program are divided by the \emph{same} constant, so as to avoid distorsive effects.}. E.g., rather than $\gamma(x_1,x_2)=N(x_1,0.1,x_2)$, we can consider $\gamma(x_1,x_2)=N(x_1,0.1,x_2)/N(0,0.1,0)$. As far as the trace semantics is concerned (Section \ref{sec:obs}), this normalization will in fact be immaterial, as it will take place on both the numerator and the denominator. On the other hand, we cannot use unbounded score functions, like say, $\gamma(x_1,x_2)=N(x_1,x_2,x_2)$.
}
\end{remark}
\fi
\vspace{-0.5cm}
\paragraph{Operational semantics of PPGs}\label{sec:PPGsem}
For any given    PPG $\ppg$, we will define a Markov kernel $\K_{\ppg}(\cdot,\cdot)$ that describes its operational semantics. From now on, we will consider one arbitrarily  fixed  PPG, $\ppg=(\St,E,\nil,\psco)$ and just drop the subscript ${}_{\ppg}$ from the notation. Let us also remark that the scoring function $\psco(\cdot)$ will play no role in the definition of the Markov kernel --- it will come into play in the trace based semantics of Section \ref{sec:obs}.

Some additional notational shorthand  is in order. First, we   identify $\PP$ with   the finite set of naturals $\{0,...,|\St|-1\}$. With this convention,  we have that  $\ereals^m\times \PP\subseteq  \ereals^{m+1}$. Henceforth, we define   our  state space and sigma-field as follows:
\begin{align*}
\Omega :=\overline\reals^{m+1} \quad \quad\quad
\extF := \text{Borel sigma-field over $ \ereals^{m+1}$.}
\end{align*}
We keep the symbol $\F_k$ for the Borel sigma-field over $\overline\reals^{k}$, for any $k\geq 1$.
% $i$-th interval of this rectangle, $1\leq i\leq m$.
%Similarly, for $1\leq i\leq m$, $A_{S,i}$ is the section of $A_S$
%Over $\Omega$, we shall often make use of $\extmu:=\mu^m \times \mu_D$, where $\mu_D$ is the discrete measure over $\reals$ defined by $\mu_D(A):=|A\cap \nat|$.
%Similarly, for $1\leq i \leq m$ and $\nu$  a measure over $\overline\reals$,  we let $\delta_v[\nu/ i]:=\delta_{v_1}\times \cdots\times \nu\times \cdots \times \delta_{v_m}$, where the $i$-th   measure is $\nu$.  %Similarly for $\extmu[\nu/ i]$.
%
For any $S\in \PP$ and $A\in \extF$, we let $A_S:=\{v\in \ereals^m\,:\,(v,S)\in A\}$ be the \emph{section} of $A$ at $S$. Note that $A_S\in\F_m$, as sections of measurable sets are measurable, see \cite[Th.2.6.2, proof(1)]{Ash}. %, in particular, if $A$ is a rectangle of  $\ereals^{m+1}$, then $A_S$ is either $\es$ or a rectangle of $\ereals^m$.
%For    $v=(v_1,...,v_m)\in \overline\reals^m$,  $r\in \ereals$ and $1\leq i \leq m$, we let $v[r@ i]:=(v_1,...,r,...,v_m)$ denote  the tuple where $v_i$ has been replaced by $r$.
%Moreover, we let $v_{\text{-}i}\in \ereals^{m-1}$ be the vector obtained from $v$ by removing its $i$-th component, and let $A_{(v_{\text{-}i},S)}=\{r\in \ereals: (v[r@i],S)\in A\}\subseteq \ereals$  be the section of $A$  at  $(v_{\text{-}i},S)$, which is a measurable set in $\F_1$.
%In particular, if $A$ is a rectangle, $A_{(v_{\text{-}i},S)}$ is either $\es$ or an interval.

\begin{definition}[PPG Markov kernel]\label{def:ppgMK}
The function $\K:\Omega\times \extF\rightarrow \reals^+$ is defined as follows, for each $\omega\in \Omega$ and $A\in \extF$:
\begin{equation}\label{eq:PPGMK}
\K(\omega)(A):= \left\{\begin{array}{ll}
	\delta_\omega(A) & \text{if $\omega\notin \ereals^m\times \PP$}\\
	\sum_{(S,\varphi,\zeta,S')\in E_S} \,\varphi(v) \cdot \zeta(v)\left(A_{S'}\right)& \text{if $\omega=(v,S)\in \ereals^m\times \PP$.}
\end{array}\right.
\end{equation}
\end{definition}

\begin{lemma}\label{lemma:MKO} %For each $\omega$, the function $\K(\omega): \R\rightarrow \ereals^+$ extends uniquely to a probability measure  $\extF\rightarrow\ereals^+$.
The function $\K$ is  a Markov kernel from $\Omega$ to $\Omega$.
\end{lemma}

%\begin{remark} Actions vs. sequential composition.\mik{To do}
%\end{remark}

\section{Trace semantics for PPGs}\label{sec:obs}
%\paragraph{Trace semantics}
%For each integer $t\geq 1$,  let  $\K^t=\K\conv\cdots \conv \K$ denote the composition of $t$ copies of $\K$. Lemma \ref{lemma:MKP} ensures that $\K^t$ is a Markov kernel: hence, for each $\omega_0\in \Omega$, $\K^t(\omega_0)$ is a probability measure over $\Omega$.
% and $A\in \extF$.
In what follows, we fix an arbitrary PPG, $\ppg=(\St,E,\nil,\psco)$  and let $\K$ denote the induced Markov kernel, as per Definition \ref{def:ppgMK}.
For any $t\geq 1$,  we call $\Omega^t$   the set of \emph{paths of length $t$}. Consider now the set of paths of infinite length,  $\Omega^\infty$, that is the set of infinite sequences $\iomega=(\omega_1,\omega_2,...)$ with $\omega_i\in\Omega$. For any   $\omega^t\in \Omega^t$ and $\iomega\in \Omega^\infty$, we identify     the pair $(\omega^t,\iomega)$ with the element of $\Omega^\infty$ in which the prefix $\omega^t$ is followed by $\iomega$.  For $t\geq 1$ and a measurable $B_t\subseteq \Omega^t$, we let $\cy{B_t}:=B_t\cdot \Omega^\infty\subseteq \Omega^\infty$ be the \emph{measurable cylinder} generated by $B_t$. We let $\Cyl$ be the minimal sigma-field over $\Omega^\infty$ generated by all measurable cylinders. Under the same assumptions of  Theorem \ref{th:prod} on the measure $\mu^1$ and on the kernels $K_2,K_3,...$ there exists a unique measure $\mu^\infty$ on $\Cyl$ such that for each $t\geq 1$ and each measurable cylinder $\cy{B_t}$, it holds that $\mu^\infty(\cy{B_t})=\mu^t(B_t)$: see \cite[Th.2.7.2]{Ash}, also   known as the \emph{Ionescu-Tulcea  theorem}. In the definition below,  we let $0=(0,...,0)$ ($m$ times) and consider $\delta_{(0,S)}$, the Dirac's measure on $\Omega$ that concentrates all the probability mass in $(0,S)$.

\begin{definition}[probability measure induced by $S$]\label{def:measPP}
Let  $S\in \PP$. For each integer $t\geq 1$, we let $\mu^t_S$ be the probability measure over $\Omega^t$ uniquely defined by Theorem \ref{th:prod}(a) by letting $\mu^1=\delta_{(0,S)}$ and $K_2=\cdots =K_t=\K$.
We let $\mu^\infty_S$ be the unique probability measure on $\Cyl$ induced by $\mu_1$ and $K_2=\cdots =K_t=\cdots=\K$, as determined by the Ionescu-Tulcea theorem.
%For each integer $t\geq 0$,  let  $\K^t$ denote the product of $t$ copies of $\K$. %Let $S\in \PP$.
%
%consider the trivial Markov kernel $i_S$ defined by $ i_S(\omega)(A)= \delta_{(0,S)}(A)$, and consider  the product Markov kernel  $i_S\conv \K^t$, defined over $\Omega^{t+1}$.
%For each integer $t\geq 1$ and $S\in \PP$, we let the \emph{probability measure at time $t$ induced by $S$} over $\Omega$   be $\nu_S^t:=\K^t(0,S)$.
%: %  Explicitly:%, for $0\in \reals^m$ and     $A\in \extF$, we have:
%\begin{align}\label{eq:nut}
%\nu^t_S(A)&:=   \K^t(0,S)(\Omega^{t-1}\times A)\,.
%\cdot \int \K(\omega_0)(d\omega_1)\cdot \;\cdots\; \cdot \int_A \K(\omega_{k-1})(d\omega_k)\,.
%\end{align}
%\ifmai
%\begin{align}\label{eq:nut}
%\nu^t_S(A)&:= \int \K(0,S)(d\omega_0)\cdot \int \K(\omega_0)(d\omega_1)\cdot \;\cdots\; \cdot \int_A \K(\omega_{k-1})(d\omega_k)\,.
%\end{align}
%$\nu^t_S$ is a probability measure on $\Omega$.
%\fi
\end{definition}

In other words, $\mu^t_S=\delta_{(0,S)}\otimes \K\otimes \cdots \otimes \K$ ($t-1$ times $\K$). By convention,  if $t=1$, $\mu^t_S=\delta_{(0,S)}$. The measure $\mu^\infty_S$ can be informally interpreted as the limit of the measures $\mu^t_S$ and represents the semantics of $S$.

Recall that the \emph{support} of an (extended) real valued function $f$ is the set $\supp(f):=\{z\,:\,f(z)\neq 0\}$. In what follows, \emph{we shall concentrate on nonnegative measurable functions} $f$ to avoid unnecessary complications with the existence of integrals. General functions can be dealt with by the usual trick of decomposing $f$ as $f=f^+ - f^-$, where $f^+=\max(0,f)$ and $f^-=-\min(0,f)$, and then dealing separately with $f^+$ and $f^-$. Let us introduce a  \emph{combined score function}   $\scoz:\Omega\rightarrow [0,1]$ as follows, for each $\omega=(v,S)$:
%$$
%\scoz(\omega):= \psco(S)(v)\,.
\begin{equation}\label{eq:scoz}
\scoz(\omega):= \left\{\begin{array}{ll}\psco(S)(v) &\text{ if $\omega=(v, S)\in \ereals^m\times \PP$}\\
1 & \text{ otherwise.}
\end{array}\right.
\end{equation}
The function $\scoz(\cdot)$ is extended to a \emph{weight function} on infinite traces, $\wt:\Omega^\infty \rightarrow [0,1]$ by letting\footnote{Note that $\wt(\tilde\omega)$ is well-defined
%by the monotone convergence theorem,
because $0\leq \scoz(\omega_j)\leq 1$ for each $j\geq0$, so the sequence of partial products is  nonincreasing.}, for any $\tilde\omega=(\omega_1,\omega_2,...)\in \Omega^\infty$:
{\small
\begin{equation}\label{eq:wt}
\wt(\tilde\omega):=\Pi_{j\geq 1}\scoz(\omega_j)\,.
\end{equation}
}\noindent
For each $t\geq 1$, we define the weight function truncated at time $t$,    $\wt_t:\Omega^t\rightarrow [0,1]$, by $\wt_t(\omega^t):=\Pi_{j=1}^t\scoz(\omega_j)$.
Both $\wt$ and $\wt_t$ ($t\geq 1$) are measurable functions  on the respective domains\iffull (see Lemma \ref{lemma:aux})\fi. We arrive at the definition of the semantics of programs. We consider the ratio of the unnormalized semantics ($[S]f$) to the weight of all traces, terminated or not ($[S]\wt$). In the special case when the score functions represent conditioning, this choice corresponds to quotienting over the probability of \emph{non failed} traces.     In PPL, quotienting over non failed states is   somewhat standard: see e.g. the discussion  in \cite[Section 8.3.2]{KaminskiTh}.

%B. Kaminski. Advanced weakest precondition calculi for probabilistic programs. 2019
%\begin{lemma}\label{lemma:measw} The functions $\wt$ and $\wt_t$ ($t\geq 1$) are measurable.
%\end{lemma}
%\begin{proof} To do.
%For each $t\geq 1$, consider the function $\tilde\wt_t$ defined by $\tilde\wt_t(\tilde\omega):=\Pi_{j=1}^t\wt(\omega_j)$. Any such
%\end{proof}
\begin{definition}[trace semantics]\label{def:obsPP}
%Let $S\in \PP$ and assume  $\mu^t_S (\term) >0$.
%Then the\emph{observable semantics at time $t$  of $S$} is given by the probability measure $\nu^t_S(\cdot|\term)$.
Let $f$ be a nonnegative measurable function defined on $\Omega^\infty$.  We let the \emph{unnormalized semantics of $S$ and $f$} be $[S] f := \expc_{\mu^\infty_S}[f]\,(=\int \mu^\infty_S(d \iomega)f(\iomega))$. %Let  the support of $f$ be contained in $\cF_\infty$, then
We let
%\vspace{-0.3cm}
\begin{align}\label{eq:semPP}
\sem{S}f&:=\dfrac{[S] (f\cdot \wt)  }{[S] \wt} %\;=\; \dfrac{\int \mu^t_S(d\omega)f(\omega)}{\mu^t_S (\term) }\,.
\end{align}
%\vspace{-0.3cm}
provided   the denominator above is $>0$; otherwise $\sem{S}f$ is undefined. %Let $\{f\}=\{f_t:\Omega^t\rightarrow \erealspl\,:\,t=1,2,...\}$ be a sequence of nonnegative measurable functions with support contained in $\NF^t$. We let $\sem{S}\{f\}:=\lim_{t\rightarrow +\infty} \sem{S}^tf_t$, if this limit exists.
\end{definition}
\vspace{-0.5cm}

\section{Finite approximation}\label{sec:FA}
The operational semantics of a probabilistic program is defined over infinite traces, due to the possibility of unbounded loops. Yet in practice,
we can reason about or sample only \emph{finite} traces. The main result of this section, Theorem \ref{th:approx}, provides a rigorous way to approximate expectations over infinite traces by computing expectations over finite prefixes.
%, while carefully accounting
%for the effect of conditioning via the weight function ??????.
Intuitively, the theorem says that if we truncate all traces at a fixed length $t$, and restrict our attention
to those that have already terminated by that point, then we can compute lower and upper bounds for
the expectation of $f$. Moreover, the bounds converge to the true value as $t\rightarrow+\infty$, if the program is guaranteed to
terminate within finite time (Theorem \ref{th:tight})

In more detail, we are interested in   $\sem{S} f$ in cases where the value of $f$  is, so to speak, determined by a finite prefix of its argument: we call these functions \emph{prefix-closed}, and will define them further below. We first have to introduce prefix-closed languages\footnote{In the context of model checking, these languages arise as complements of Safety properties; see e.g. \cite[Def.3.22]{BaierKatoen}.}, for which   some notation on languages of finite and infinite words is useful. Given two words $w,w'\in \Omega^*$, % and $w'\in \Omega^*\cup \Omega^{\infty}$,
we write $w\prec w'$ if $w$ is a prefix of $w'$, i.e. there exists a word $w''\in \Omega^*$ such that $ww''=w'$; otherwise we write $w\not\prec w'$. For $L,L'\subseteq \Omega^*$, we write $L\not\prec L'$ if for all $w\in L$ and $w'\in L'$ we have $w\not\prec w'$. A sequence of languages $L_0,L_1,...$ such that for each $j$, $L_j\subseteq \Omega^j$ (with $\Omega^0:=\{\epsilon\}$, the empty sequence) is said to be \emph{prefix-free} if for each $i\neq j$, $L_i\not\prec L_j$. Note that if $L_0\neq \es$ then $L_j=\es$ for $j\geq 1$. For the sake of uniform notation, in what follows we convene that    $\omega^0:=\epsilon$ and $\cy{\{\epsilon\}}:=\Omega^\infty$. We say   $A\subseteq \Omega^\infty$ is a \emph{prefix closed set} if there is a prefix-free sequence of languages $L_0,L_1,...$ such that $A=\cup_{j=0}^\infty \cy{L_j}$; we call   $L_j$   a \emph{$j$-branch} of $A$, and refer to $L_0,L_1,...$ collectively as   \emph{branches of} $A$.  For   any $t\geq 1$,  we define the following subsets of $\Omega^t$:
\begin{align*}
L^{\leq t}&:=\cup_{j=0}^t L_j\cdot \Omega^{t-j}, \ \ \ \ \ \ \ \ \ L^{> t}:=\{\omega^t\,:\, \text{ there is $t'>t$ and $\omega_{t'}\in L_{t'}$ s.t. } \omega^t \prec \omega^{t'}  \}\,.
\end{align*}
Informally speaking,   $L^{\leq t}$ is the set of paths of length $t$ that will become members of $A$ however we extend them to infinite words. $L^{> t}$ is  the set of  paths of length $t$ for which some infinite extensions, but not all, are in  $A$ --- they are so to speak   “undecided''.  Of special interest is the     prefix-free sequence of languages defined below.

\begin{definition}[termination]\label{def:term}
Let $\term:=\ereals^m\times \{\nil\}$ be the set of \emph{terminated} states and let $\term^{\mathrm{c}}$ denote its complement. We let  $T_j\subseteq\Omega^j$ ($j\geq 0$) be the set of finite sequences that \emph{terminate at time} $j$, that is: $T_0:=\es$ and   $T_j:=(\term^{\mathrm{c}})^{j-1}\cdot \term$, for $j\geq 1$. We let $T_{\fini}:=\cup_{t\geq 0}\cy{T_t}\subseteq \Omega^\infty$ denote the set of infinite sequences that \emph{terminate in finite time}.
%We say a prefix closed language $A$ with branches $L_j$ ($j\geq 0$) is \emph{T-respectful} if for each $j\geq 0$, $L^{> j}\cap T_j=\es$.
\end{definition}

Note that $\{T_j:j\geq 0\}$  forms a prefix-free sequence,   that $T^{\leq t}\subseteq \Omega^t$ is the set of all paths of length $t$  that terminate within time $t$, while     $\cy{T_t}\subseteq\Omega^\infty$ is the set of  infinite execution paths with termination at time $t$.
%, while $\term^{\mathrm{c}}:=\Omega\setminus \term$ is the set of non-terminated states\footnote{It is immaterial
% to consider junk states $(v,z)$ with $z\notin \PP$  as non-terminated.}
%In particular, .
%, while $T^{> t}$ is the set of all paths of length $t$ that do not terminate within time $t$. % \emph{and} can be extended to a terminated path of length  $t'>t$.
The next definition introduces   prefix-closed    functions. These are functions $f$ with a prefix-free support, condition (a), additionally satisfying two extra conditions. Condition (b) just states that the value of $f$ on its support is determined by a finite prefix of the input sequence.
Condition (c), T-respectfulness,   means that  a trace that terminates at time   $j$ ($\omega^j\in T_j$) cannot lead to  $\supp(f)$ at a later time ($\omega^j\notin L^{> j}$).
%, formalizes that  finite execution paths terminating  at time $j$  ($\omega^j\in T_j$) cannot be undecided  w.r.t. the prefix-closed support of $f$ (hence $\omega^j\notin L^{>j}$). When required to hold true for all $j$,
This is a consistency condition,  formalizing that  the value of $f$ does not depend on, so to speak, what happens \emph{after} termination.
%We also define \emph{lifting},  a natural way of defining termination based functions from functions on $\Omega$ that only look at the value of variables in (correctly) terminated states.
%We are mainly interested in the $\sem{S} f$ when $f$ is termination based.

\begin{definition}[prefix-closed function]\label{def:termf}
Let $f:\Omega^\infty \rightarrow\erealspl$ be a nonnegative measurable function and $(L_0,L_1,...)$ be a prefix-closed sequence. We say $f$ is a \emph{prefix-closed function} with branches $L_0,L_1,...$ if the following conditions are satisfied.
\begin{itemize}
\item[(a)] $\supp(f)$ is   prefix-free with   branches $L_j$ ($j\geq 0$).
\item[(b)] for each $j\geq 0$ and $\omega^j\in L_j$, $f$ is constant on $\cy{\{\omega^j\}}$.
\item[(c)] $\supp(f)$ is \emph{T-respectful}:   for each $j\geq 0$, $L^{> j}\cap T_j=\es$.
\end{itemize}
%and for each $t\geq 1$ and  $\omega^t\in \live^{t-1}\times \term$, we have that $f$ is constant on $\{\omega^t\}\times \Omega^\infty$.
%
%	For any such $f$ and $t\geq 1$, we let $f_t:\Omega^t\rightarrow \erealspl$ be defined by $f_t(\omega^t):=f(\omega^t,*^\infty)$, for an arbitrarily fixed $*\in\live$.
%	
%Given a nonnegative measurable $g$ on $\Omega$ with  $\supp(g)\subseteq \ereals^m\times \{\nil\}$, we define the \emph{lifting} of $g$ to   $\Omega^\infty$  as the function $f$   defined as follows: for $\iomega=(\omega_1,\omega_2,...)$,  $f(\iomega):=g(\omega_t)$ if  $\iomega\in \term_t$ for some $t\geq 1$; otherwise, $f(\iomega):=0$. The lifting of $g$ to $\Omega^t$ is $f_t$, where $f$ is the lifting of $g$ to $\Omega^\infty$.
\end{definition}

Note that there may be different prefix-free sequences w.r.t. which $f$ is prefix-closed.

%Most functions of interest can be defined via lifting.

%In what follows, we let $\term:=\ereals^m\times \{\nil\}$ be the set of \emph{terminated} states, while $\term^{\mathrm{c}}:=\Omega\setminus \term$ is the set of non-terminated states\footnote{It is immaterial
% to consider junk states $(v,z)$ with $z\notin \PP$  as non-terminated.}.

%\mik{Aggiungere esempi}
\begin{example}{%[prefix-closed functions]%[terminating paths]
\label{ex:f}
%Let us define the following prefix-free sequence of languages: $T_0:=\es$ and   $T_t:=(\term^{\mathrm{c}})^{t-1}\cdot \term$ for $t\geq 1$. Then $\cy{T_t}$ is the set of execution paths that terminate at time $t$.
%The sequence $T_0,T_1,T_2,...$ is clearly prefix-free.
%The language $T_{\fini}:=\cup_{t\geq 0}\cy{T_t}$ is the set of   execution paths that terminate in finite time, and is a prefix closed set.
The indicator function  $1_{T_{\fini}}$ is clearly a prefix closed, measurable function with $\supp(1_{T_{\fini}})=T_{\fini}$ and branches $L_j=T_j$. For more interesting examples, consider the PPG in Example \ref{ex:dm1} and the functions
%\begin{itemize}
%	\item $f_1(\iomega)=j$ if $\omega_j\in \term$ is the first terminated state occurring in $\iomega$, if   such a $\omega_j$ exists; 0 otherwise.
%	\item $f_2(\iomega)=d$ if $\omega=(d,r,x,y,\nil)\in \term$ is the first terminated state occurring in $\iomega$ and $d\in [0,2]$, if   such a $\omega_j$ exists; 0 otherwise.
%	%\item $f_2(\iomega)=1$ if there are $\omega_j=(d,r,x,y,S)$ and $\omega_{j+1}=(d',r',x',y',S')$ occurring in $\iomega$ such that   and $|y-y'|> 3$, and $\nil$ does not occur in any   state preceding $\omega_k$ in $\tilde\omega$; 0 otherwise.
%	
%	%\item $f_4(\tilde\omega)=1$ if there are $\omega_i=(r,y,i,S)$ and $\omega_{j}=(r',y',i',S')$ occurring in $\tilde\omega$ ($i\neq j$) such that   $|y-y'|<0.01$; 0 otherwise.
%\end{itemize}
$f_1$, that returns  the time the process terminates, and  $f_2$ that returns the value of $d$ at termination. Here $\supp(f_1)=T_{\fini}$ has   branches $L_{j}=T_j$ ($j\geq 0$), instead $\supp(f_2)=\{\iomega\in T_{\fini}\,:\,$ the first terminated state $\omega$ in $\iomega$, if it exists, has   $\omega(1)=d\in(0,2]\}$, and $L_0=\es$, $L_j=(\term^{\mathrm{c}})^{j-1}\cdot (\term\cap ((0,2]\times \ereals^4))$ ($j\geq 1$).
%, where $A=\{(d,r,x,y,z):  |x-y|\geq 1/10  \text{ and }z\neq \nil\}$ and $A=\{(d,r,x,y,z):  |x-y|< 1/10  \}$.  The condition about $\nil$ ensures the T-respectfulness of the prefix-free sequence of the $L_j$s: indeed $\omega^j\in \term_j$ implies $\omega^j\notin L^{>j}$.
%The function $f_2$ indicates that,   at some point    before termination, the mouse takes a transition with an exceptionally long step ($>3$ in s.d. units).

% such that  \omega_j=(d,r,x,y,S) \text{ there is   $\omega_i=(d,r,x,y,S)$ s.t. }   obtained by replacing $\iomega$ with $\omega^j$ in the condition for 1 in its definition.
%$L_j=\{\omega^{j}\,:\,$The branches of $f_2,f_3$ can be worked out similarly.
}\end{example}

%Given a termination based  $f$ and $t\geq 1$, let us define $f_t:\Omega^t\rightarrow \erealspl$ by $f_t(\omega^t):=f(\omega^t,*^\infty)$, for an arbitrarily fixed $*\in\live$.

We will now study how to consistently approximate infinite computations ($\mu^\infty_S$ semantics) with finite ones ($\mu^t_S$ semantics). This will lead to the main result of this section (Theorem \ref{th:approx}). As a first  step, let us   introduce an appropriate notion of finite approximation for functions $f$ defined on the infinite product space $\Omega^\infty$.   Fix an arbitrary element $\star\in \Omega$. For each $f:\Omega^\infty \rightarrow \ereals^+$ and $t\geq 1$, let us define the function $f_t:\Omega^t \rightarrow \ereals^+$ by $f_t(\omega^t):=f(\omega^t,\star^\infty)$.
The intuition here is that, for a prefix-closed function $f$, the function  $f_t$ approximates correctly $f$ for all  finite paths in the $L_j$-branches of $f$, for $j\leq t$. Consider for instance the function $f=f_1$ in Example \ref{ex:f}. On $L^{\leq t}$, the approximation $f_t$ gives the correct value w.r.t. $f$ in a precise sense: $f_t(\omega^t)=f(\omega^t,\star^\infty)=f(\omega^t,\tilde\omega')$  whatever   $\star$ and $\tilde\omega'$. On the other hand, for finite paths $\omega^t\in L^{>t}$, $f_t$ may not approximate $f$ correctly: we may have  $f_t(\omega^t)=f(\omega^t,\star^\infty)\neq f(\omega^t,\tilde\omega')$ depending  on the specific $\star$ and $\tilde\omega'$. The catch is, as $t$ grows large, the set $L^{>t}$ will become thinner and thinner --- at least under   reasonable assumptions on the measure $\mu^\infty_S$.

It is not difficult to check that, for any $t$,  $f_t$ is measurable over $\Omega^t$\iffull (Lemma \ref{lemma:aux} in the Appendix).\else.\fi The next result shows how to approximate   $\sem S f$ with quantities defined \emph{only in terms of $f_t, \wt_t$ and $\mu^t_S$}, which is the basis for the   sampling-based inference algorithm  in the next section.
Formally, for $t\geq 1$ and a   measurable function $h:\Omega^t\rightarrow \ereals^+$, we let
$$%\begin{equation*}
\begin{array}{ll}
[S]^t h&:=\expc_{\mu^t_S}[h] \,\,(=\int \mu^t_S(d\omega^t)h(\omega^t)\,)\,.
\end{array}
$$%\end{equation*}
%
%
%Clearly, $\Omega^t=\term^t\cup \failE^t\cup \live^t$. We let the set of \emph{non failed} paths of length $t$ be $\NF^t:=\term^t\cup \live^t$.
%Note that $\term^t\subseteq \live^t$, that $\live^t\cap \failE^t=\es$ and that $\Omega^t=\live^t\cup \failE^t$.
%We may omit the superscript $t$   when $t$ is clear from the context. %Let $\term^t:=\Omega^t\setminus(\reals^m\times \{\fail\})^t$ represent the \emph{non-failure} event in $\extF^t$.
%
%
%
%
The intuition of the theorem is as follows. Consider a prefix closed function $f$ with branches $L_0,L_1,...$.
For  any time   $t$,  it is not difficult to see that   $\cy{{L}^{\leq t}\cap T^{\leq t}} \subseteq \supp(f)\subseteq \cy{{L}^{\leq t}\cap T^{\leq t}}\cup (\cy{{T}^{\leq t}})^{\mathrm{c}}$ (the last inclusion involves T-respectfulness).
%The finite approximation theorem for $\sem S f$ relies on the observation that,
Since $f_t$ approximates correctly $f$ on $L^{\leq t}$, one sees that the first inclusion leads to the lower bound  $[S]^t f_t\cdot 1_{L^{\leq t}\cap T^{\leq t}}\cdot \wt_t\leq [S]f\wt$. As for the upper bound, the intuition is that, over $(\cy{{T}^{\leq t}})^{\mathrm{c}}$, $f$ is upper-bounded by $M$.

\begin{theorem}[finite approximation]\label{th:approx} Consider $S\in\St$ and $t\geq 1$ such that $[S]^t 1_{T^{\leq t}}\cdot \wt_t>0$. %, $\mu^t_S(T^{\leq t})>0$.
Then for any  prefix-closed   function $f$ with branches $L_0,L_1,...$ we have that $\sem{S}f$ is well defined. % and finite. %, albeit possibly $=+\infty$.
Moreover,  given  an upper bound  $f\leq M$ ($M\in \erealspl$),    for each $t$ large enough and $\alpha_t:=\frac{[S]^t \wt_t}{[S]^t 1_{T^{\leq t}}\cdot \wt_t}$ we have:
%\begin{itemize}
%\item[(a)] Given  an upper bound  $f\leq M$ ($M\in \erealspl$),    for each $t$ large enough:
\begin{equation}\label{eq:approx}
\begin{array}{rcccl}
	\dfrac{[S]^t f_t\cdot 1_{L^{\leq t}\cap T^{\leq t}}\cdot \wt_t}{[S]^t \wt_t} &\leq & \sem{S}f&
	\leq &
	\dfrac{[S]^t f_t\cdot 1_{L^{\leq t}\cap T^{\leq t}}\cdot \wt_t}{[S]^t \wt_t}\alpha_t+M\cdot   \left(\alpha_t   -1\right)\,.
	%\dfrac{[S]^t  (f_t\cdot  1_{L^{\leq t}\cap T^{\leq t}}+M\cdot   1_{L^{>t}}) \cdot \wt_t }{[S]^t 1_{T^{\leq t}}\cdot \wt_t}\,.
\end{array}
\end{equation}
%where $L_0,L_1,...$ are  the branches of $f$.
\end{theorem}

When $f$ is an indicator function, $f=1_A$,   we can of course take $M=1$ in the theorem above. We first illustrate the above result with a simple example.

\begin{example}\label{ex:simple1}
Consider the PPG of Example \ref{ex:simple0} (Fig. \ref{fig:drunk}, left).
We ask what is the expected value of $c$ upon termination of this program. Formally, we consider the program checkpoint $S=0$, and the function $f$  on traces that returns the value of $c$ on the first terminated state, if any, and 0 elsewhere. This $f$ is clearly prefix-closed with branches $L_j\subseteq T_j$. We apply Theorem \ref{th:approx} to $\sem S f$. Fixing the time $t = 4$, we can calculate   easily the quantities involved in the approximation of $\sem S f$ in \eqref{eq:approx}. In doing so, we must consider  the finitely many paths of length $t$ of nonzero probability and weight (there are only two of them),  their weights and the value of $c$ on their final state when terminated\footnote{Here,  we also use the fact that $f_t\cdot 1_{T^{\leq t}}=f_t\cdot 1_{L^{\leq t}\cap T^{\leq t}}$, a consequence of $L_j\subseteq T_j$ for all $j$s.}.
%{\small
%\begin{align*}
{\small
	%\begin{itemize}
	$$
	\begin{array}{ll}
		[S]^t f_t\cdot 1_{T^{\leq t}}\cdot \wt_t =0\cdot \frac 1 2 + 1\cdot \frac 1 2\cdot \frac 1 2=  \frac{1}{4}
		\hspace*{.5cm}&
		[S]^t  \wt_t = \frac 1 2   +  \frac 1 2\cdot \frac 1 2 = \frac{3}{4} \\  %
		{[}S{]}^t  1_{T^{\leq t}}\cdot\wt_t =\frac 1 2   +  \frac 1 2\cdot \frac 1 2 = \frac{3}{4} & \alpha_t   = 1\,.
		%\end{align*}
	\end{array}
	$$
	%\end{itemize}
}\noindent
Then, with $M=1$,  the lower and upper bounds in \eqref{eq:approx} coincide and yield $ \sem S f = \frac 1 3$.
% $1.94\cdots=\frac{37}{19}\leq \sem S f\leq \frac{37}{19}+\frac{30}{19}=3$. Actually, in this case the exact value coincides with the lower bound, $\sem S f=\frac 2 5$.
If we remove conditioning on node 4, then all the paths of length $t$ have weight 1, and a similar calculation yields   $ \sem S f= \frac 1 2$.
%$\frac 4 3 \leq \sem S f\leq 1$, with again the lower bound being the exact value.
\end{example}

In more complicated cases, we may not be able to calculate exactly the quantities involved in \eqref{eq:approx}, but only to estimate them via sampling. To this purpose, we will introduce Feynman-Kac models and the Particle Filtering algorithm in the next section. %For now, we content ourselves with the following example.
%
%\begin{example}{\label{ex:RW2}
%	%\mik{Completare esempio}
%	Consider the PPG of Example \ref{ex:dm1} and   $f=f_2$ from Example \ref{ex:f}.  Take $S=0$. Then $\sem{S}f$ is the posterior expectation of the value of $d$, the drunk man's standard deviation. We can compute upper and lower bounds on  $\sem{S}f$ using \eqref{eq:approx}. Let us fix $t=60$. By sampling from $\mu^t_S$, we can compute separately the following estimates for each of the expected values  involved in  \eqref{eq:approx}:
%	$[S]^t f_t\cdot 1_{L^{\leq t}\cap T^{\leq t}}\cdot \wt_t=  0.396 $, $[S]^t \wt_t=0.987$ and $\alpha_t=1.497$.
%	%$[S]^t 1_{T^{\leq t}}\cdot \wt_t=0.17668$,
%	Combining these estimates as in \eqref{eq:approx}, with $M=2$, we get the bounds: $0.402\leq  \sem{S}f\leq 1.596$. This   relatively large interval can be narrowed down by considering higher values of $t$, hence $\alpha_t$ closer to 1.
%	%For comparison, in a random walk model with only the drunk man   (and no scoring function), one gets the bounds:  $0.9989\leq  \sem{S}f\leq3.536$.
%	A more efficient and accurate  method to compute the bounds in \eqref{eq:approx} will be introduced in Section \ref{sec:MC}, the Particle Filtering algorithm.
%}\end{example}
The theorem below confirms that the bounds established above are asymptotically tight, at least under the assumption that the program $S\in \PP$ terminates with probability 1. In this case, in fact, the probability mass outside $\term^{\leq t}$ tends to 0, which leads the lower and the upper bound in \eqref{eq:approx} to coincide. Moreover, we get a simpler formula in the special case when termination is guaranteed to happen within a fixed time limit; for instance, in the case of    acyclic\footnote{Or, more accurately, PPGs where the only loop is the self-loop on the \textsf{nil} state.} PPGs.

%\begin{definition}[T-respectfulness]\label{def:Tresp}
%A prefix closed language  with branches $L_0,L_1,...$ is \emph{T-respectful} if for every $j\geq 0$, $T_j\cap L^{>j}=\es$. We say that a prefix closed function $f$ is T-respectful if $\supp(f)$ is T-respectful.
%\end{definition}

\begin{theorem}[tightness]\label{th:tight} Assume the same hypotheses as in Theorem \ref{th:approx}. Further
assume that $\mu^\infty_S(T_\fini)=1$. % and that $f$ is T-respectful.
Then both the lower and the upper bounds in \eqref{eq:approx} tend to $\sem S f$ as $t\rightarrow +\infty$. In particular, if for some $t\geq 1$ we have    $[S]^t 1_{T^{\leq t}}=1$, % (that is $\mu^t_S(\term^t)=\mu^t_S(\live^t)$),
then
\begin{equation}\label{eq:exact}
	\begin{array}{rcl}
		\sem{S} f  &= & \dfrac{[S]^t f_t\cdot \wt_t}{[S]^t    \wt_t}\,.
	\end{array}
\end{equation}

%In particular, if for some $t$ we have $L^{\leq t}\subseteq T^{\leq t}$ and  $[S]^t 1_{T^{\leq t}}=1$, % (that is $\mu^t_S(\term^t)=\mu^t_S(\live^t)$),
%	then
%	\begin{equation}\label{eq:exact}
	%		\begin{array}{rcl}
		%			\sem{S} f  &= & \dfrac{[S]^t f_t\cdot \wt_t}{[S]^t    \wt_t}\,.
		%		\end{array}
	%	\end{equation}
%\end{itemize}
\end{theorem}

\begin{example}\label{ex:simple2} For the PPG of Example \ref{ex:simple0} one has $\mu^\infty_S(T_\fini)=1$. As already seen in Example \ref{ex:simple0}, lower and upper bounds coincide for $t\geq 4$.
\end{example}

%\begin{remark}{\em\label{rem:Tresp} The proof of the Theorem \ref{th:tight} shows that convergence of the lower bound holds even in the absence of T-respectfulness.}
%\end{remark}

A practically relevant class of closed prefix functions are those where the result    $f(\iomega)$ only depends on computing a function $h$, defined on $\Omega$, on the first terminated state, if any, of the sequence   $\iomega$.  % that sdefined via the following construction.
%For $\iomega=(\omega_1,\omega_2,...)\in \Omega^\infty$ and $j\geq 1$, let $\iomega(j)=\omega_j$.
This way  $h$ is    {\emph{lifted}} to $\Omega^\infty$. This case covers all the examples seen so far.  We formally introduce lifting below. Recall that for $t\geq 1$, $T_t=(\term^{\mathrm{c}})^{t-1}\cdot\term$.

\begin{definition}[lifting]\label{def:simple} Let $h:\Omega \rightarrow \ereals^+$ a nonnegative measurable function such that $\supp(h)\subseteq \term$. The \emph{lifting of}  $h$   is the measurable function $\lift h:\Omega^\infty\rightarrow \ereals^+$ defined as follows for each $\iomega=(\omega_1,\omega_2,...)$: $ \lift  h(\iomega):=\sum_{t\geq 1}1_{\cy{T_t}}(\iomega)\cdot h(\omega_t)$.
%A prefix closed function $f$ is \emph{simple} if there is a measurable function $h:\Omega\rightarrow \ereals^+$ such that  for each $t\geq 0$ and  $\iomega\in \cy{L_t}$, $f(\iomega)=h(\iomega(t))$, where $L_t$ is the $t$-branch of $f$.
\end{definition}

Clearly, any $\lift h$ is prefix closed with branches $L_0=\es$ and $L_j=(\term^{\mathrm{c}})^{j-1}\cdot\supp(h)\subseteq T_j$ for $j\geq 1$. In particular, $\supp(\lift h)\subseteq T_{\fini}$. As an example, the indicator function for the set of paths that eventually terminate, $\lift h=1_{\term_\mathrm{f}}$, is clearly the lifting of $h=1_{\term}$; the functions $f_1,f_2$ in   Example \ref{ex:f} can also be obtained by lifting (details omitted).

\ifmai
\begin{example}{\label{ex:lift}
	The indicator function for the set of paths that eventually terminate, $\lift h=1_{\term_\mathrm{f}}$, is clearly the lifting of $h=1_{\term}$. For more interesting examples, consider Example \ref{ex:f}.    The function $f_1$ is not a lifting of any function on $\Omega$, as it must count how many states are traversed up to  the first one in $\term$. However, one can modify the program of Example \ref{ex:dm1} to insert a new counter variable $j$  that provides this information, and return the value of $j$ at termination. The function $f_2$ is the lifting of the function $h(d,r,x,y,z)=d\cdot [0\leq d\leq 2]\cdot [z=\nil]$.

	%that gets updated in the self-loop of checkpoint 1, that is we add   $j:= j+1 $ to the self loop. In the new PPG,  one   considers the lifting of the function $h$ that, on terminated states, returns $j+2$: this $\hat h$ is (pragmatically) equivalent\footnote{Here the +2 accounts for the fact that states are numbered from 1, and for the transition that leads to $\nil$.}  to $f_1$.
	%One can encode $f_2$ with a similar   program transformation.
}\end{example}
\fi

%and nonnegative measurable function $h$ induce a unique simple function

\section{Feynman-Kac models}\label{sec:MC}
In the field of Sequential Monte Carlo methods, Feynman-Kac (FK) models \cite[Ch.9]{SMC}  are  characterized by the use of \emph{potential} functions.
A potential
in a Feynman-Kac model is a function that assigns a weight $G_t(x)$ to a \emph{particle} (instance of a random process) in state
$x$ at time
$t$. This weight represents how plausible or fit
$x$ is at time
$t$ based on some observable or conditioning. In other words,   $G_t$ modifies the \emph{importance} of particles as the system evolves. For instance, in a model for tracking an object, the potential function could depend on the distance between the predicted particle position and the actual observed position. Particles closer to the observed position get higher weights.
%Intuitively speaking,  potentials can be seen as applying a weight to states. this influence can either attract or repel the process from certain regions in the state space. As such, the use of potentials is analogous to, and in fact generalizes,     conditioning in probabilistic programming.

\vspace{-0.5cm}
\paragraph{FK models and probabilistic program semantics}\label{sub:genrel}
As seen, our semantics incorporates conditioning via score functions applied at program checkpoints, and
aggregates their effect into a global weight $w$ over traces. This makes it possible to interpret program  semantics
  as a reweighted expectation (Definition \ref{def:obsPP}). Here we will show that this expectation can be approximated reliably using the Feynman-Kac framework and particle filtering.
We first introduce FK models in a general context.  Our formulation follows closely \cite[Ch.9]{SMC}. Throughout this and the next section, we let $t\geq 1$  be an arbitrary fixed integer.

\begin{definition}[Feynman-Kac models]\label{def:FK}  A \emph{Feynman-Kac (FK) model} is a tuple $\FC=(\X,t,\mu^1,\{K_i\}_{i=2}^t,\{G_i\}_{i=1}^t)$, where $\X=\ereals^\ell$ for some $\ell\geq 1$, $\mu^1$ is a probability measure on $\X$ and, for $i=2,...,t$:  $K_i$ is a Markov kernel from $\X$ to $\X$, and $G_i:\X\rightarrow \ereals^+$ is a measurable function.

Let  $\mu^t$ denote the unique product measure on $\X^t$ induced by $\mu_1,K_2,...,K_t$ as per Theorem \ref{th:prod}. Let $G:=\Pi_{i=1}^t G_i$. % and $P_t:=\expc_{\mu^t}[G]$.
Provided $0<\expc_{\mu^t}[G]<+\infty$, the \emph{Feynman-Kac measure} induced by $\FC$ is defined by the following, for every measurable $A\subseteq \X^t$:
	\vspace{-0.2cm}
\begin{align}\label{eq:FCM}
	\mufk_{\FC}(A)&:=\frac{\expc_{\mu^t}[1_A\cdot  G]}{\expc_{\mu^t}[G]}\,.
\end{align}
%which is defined provided
\end{definition}
	\vspace{-0.1cm}
%The functions  $G_i$ are also  called  \emph{potential} at time $i$;
We will refer to $G$ in the above definition as the \emph{global potential}.
Equality \eqref{eq:FCM} easily generalizes to expectations   taken according to $\mufk_{\FC}$. That is, for any measurable nonnegative function $g$   on $\X^t$, we can easily show that:
	\vspace{-0.3cm}
\begin{align}\label{eq:FCMe}
\expc_{\mufk_{\FC}}[g]&=\frac{\expc_{\mu^t}[g\cdot  G]}{\expc_{\mu^t}[G]}\,.
	\vspace{-0.2cm}
\end{align}
In what follows, we will suppress  the subscript ${}_{\FC}$ from ${\mufk}_{\FC}$ in the notation,  when no confusion   arises. Comparing  \eqref{eq:FCMe} against the definition \eqref{eq:semPP} suggests that the global potential $G$ should play in FK models a role analogous to the weight function $\wt$ in probabilistic programs. Note however that there is a major technical difference between the two, because FK models are only defined for a finite time horizon model given by $t$. A reconciliation between the two is possible thanks to the finite approximation theorem seen in the last section; this will be elaborated further below (see Theorem \ref{th:filtlift}).
%there is e.g. the lower bound in  \eqref{eq:approx} suggests   considering the following FK model associated with $\ppg$ and a checkpoint $S$.

We will be  particularly interested in the \emph{$t$-th marginal} of $\mufk$, that is the probability measure on $\X$ defined as ($A\subseteq \X$ measurable):
\begin{equation}\label{eq:phit}
\mufk_t(A):=\mufk(\X^{t-1}\times A)  =  \expc_\mufk[1_{\X^{t-1}\times A}]\,.
\end{equation}
The measure $\phi_t$ is called \emph{filtering} distribution (at time $t$), and can be effectively computed via the Particle Filtering algorithm   described in the next subsection.
Now  let $\ppg=(\St,E,\nil,\psco)$ be an arbitrary fixed PPG.  Comparing  \eqref{eq:FCMe} against e.g. the lower bound in  \eqref{eq:approx} suggests   considering the following FK model associated with $\ppg$ and a checkpoint $S$.
%a  with $\X=\Omega$, $\mu_1=\delta_{(0,S)}$,  $K_i=\K$  and $G_i=\scoz$, hence $G=\wt_t$. However, this is adequate only in the exact case in which \eqref{eq:exact} holds (termination within time $t$). In the more general case, we have to take into account the possibility of non-termination within time $t$. To any   program $S\in \St$ we will associate \emph{two} distinct FK
%models, hence measures,  $\FC^{-t}_S$ and  $\FC^{+t}_S$, that correspond to the lower bound and the upper bounds in \eqref{eq:approx}.  $\FC^{-t}_S$ is just the FK model outlined above, while in $\FC^{+t}_S$   particles not terminated at time $t$ will be assigned 0 weight. In fact, the two models only differ as far as the  potential function at time $t$,  $G_t$, is concerned. Formally, we have the following.

\begin{definition}[$\FC_S$  model]\label{def:PPFK} Let $t\geq 1$ be an integer and $S$ a program checkpoint of $\ppg$. We define $\FC_S$ as the FK model where: $\X=\Omega$, $\mu^1=\delta_{(0,S)}$, $K_i=\K$ ($i=2,...,t$) and $G_i=\scoz$ ($i=1,...,t$).
We let $\phi_S$  denote the measure on $\Omega^t$ induced by $\FC_S$.
\end{definition}

%The extension of a fixed-horizon FK model to an infinite horizon is given in \cite{delmoral2}. \lui{aggiunto rif. Del Moral}
We now restrict our attention to functions $f$ that are the lifting of a nonnegative $h$ defined on $\Omega$.
Let $\phi_{S,t}$ denote the filtering distribution of $\phi_{S}$ at time $t$ obtained by \eqref{eq:phit}. In the following theorem we express the bounds in \eqref{eq:approx} in terms of the measure  $\phi_{S,t}$. The whole point and interest of this result is that the bounds are expressed  directly as expectations; these are moreover taken w.r.t. a  \emph{1-dimensional} filtering distribution ($\phi_{S,t}$),  rather than  a   $t$-dimensional one ($\mu^t_S$). Importantly, there are well-known algorithms to estimate expectations under a filtering distribution, as we will see in the next subsection.

%\mik{Completare.}
\begin{theorem}[filtering distributions and lifted functions]\label{th:filtlift} Under the same assumptions of Theorem \ref{th:approx}, further assume that $f$ is the lifting of $h$. Then $\alpha_t= \expc_{\phi_{S,t}}[1_{\term}]^{-1}$ and
\begin{equation}\label{eq:filtering}
	\begin{array}{rcl}
		\beta_L:=\;\expc_{\phi_{S,t}}[h] &\leq  \,\, \sem S f \,\,\leq &  \expc_{\phi_{S,t}}[h]\cdot \alpha_t \;+\;M\cdot (\alpha_t-1)\;=:\beta_U\,.
	\end{array}
\end{equation}
\end{theorem}

\begin{example}\label{ex:simple3}
Consider again the PPG of Example \ref{ex:simple0}. We can re-compute $\sem S f$ relying on Theorem \ref{th:filtlift}. Fix $t=4$. We first compute the filtering distribution $\phi_t$ on $\X=\Omega=\ereals^3$ relying on its definition \eqref{eq:phit}. Similarly to what we did in Example \ref{ex:simple1}, we   consider the nonzero-weight, nonzero-probability traces of length four. Then we project onto   the final (fourth) state, and compute the weights of  the resulting triples $(c,d,S)$, then normalize. There are only two triples $(c,d,S)$ of nonzero probability: $\phi_t(0,0,2)=\frac 2 3$, $\phi_t(1,1,2)=\frac 1 3$.
%\vspace*{-0.2cm}
%{\small
%	\begin{align*}
%		\phi_t(0,0,2)&=\frac 2 3 &
%		\phi_t(1,1,2)&=\frac 1 3\,.
%		\vspace*{-0.2cm}
%	\end{align*}
%}\noindent
The function $f$ considered in Example \ref{ex:simple1} is the lifting of the function   $h(c,d,S)=c\cdot  {[S=2]}$ defined on $\X=\ereals^3$. We apply Theorem \ref{th:filtlift} and get $\beta_L= \expc_{\phi_t}[h]=\frac 1 3\leq \sem S f$. Moreover $\expc_{\phi_t}[1_\term] =1$, hence $\alpha_t=1$ according to Theorem \ref{th:filtlift}. Hence $\beta_L=\beta_U=\sem S f = \frac 1 3$. This agrees with   examples   \ref{ex:simple1} and   \ref{ex:simple2}.
\end{example}

%\noindent
We can apply the above theorem to the functions described in %Example \ref{ex:f}, either directly or via the encodings outlined in
Example \ref{ex:f} and to other computationally challenging cases: we will do so in Section \ref{sec:experiments}, after introducing in the next section the Particle Filtering algorithm.
%For the time being, we shall illustrate Theorem \ref{th:filtlift} via a simpler example.

%\begin{example}\label{ex:simpler}
%To do.
%\end{example}
%\vspace*{-.2cm}
%In what follows, for any given a $S\in \P$, we introduce three FK models. Our objective is used the corresponding measures to express the bounds in \eqref{eq:approx} and the (exact) expression \eqref{eq:exact}.
\vspace{-0.4cm}
\paragraph{The Particle Filtering algorithm}\label{sub:algo}
From a computational point of view, our interest in FK models lies in the fact that they allow for a simple, unified presentation of a class of efficient inference algorithms,  known as \emph{Particle Filtering (PF)} \cite{SMC,DelMoral04,wood}. \lui{aggiunto rif. Wood}   For the sake of presentation,  we only introduce here the basic  version, \emph{Bootstrap} PF,  following closely\footnote{\iffull Additional details in Appendix  \ref{app:PF}.\else Additional details in \cite{BC25}.\fi} \cite[Ch.11]{SMC}.
%In what follows we present an algorithm to compute the filtering distribution $\phi_t$. We will introduce below a general PF algorithm scheme  following closely .
%
Fix  a generic FK model, $\FC=(\X,t,\mu^1,\{K_i\}_{i=2}^t,\{G_i\}_{i=1}^t)$. Fix  $N\geq 1$, the number of \emph{particles}, that is instances of the random process represented by the $K_i$'s, we want to simulate.  Let  $W=W^{1:N}=(W^{(1)},...,W^{(N)})$   be a tuple of $N$ real nonnegative random variables, the  \emph{weights}.
Denote by $\widehat W$ the normalized version of $W$, that is
%\footnote{With the proviso that e.g. $\widehat W_i:=1/N$ in the event all the $W_i$'s are 0.  In the actual execution of the PF algorithm this event will occur with   probability $\rightarrow 0$ as $N\rightarrow +\infty$.}
$
\widehat W^{(i)}=W^{(i)}/(\sum_{j=1}^N W^{(j)})$.
A \emph{resampling scheme} for  $(N,W)$ is a $N$-tuple of random variables $R=(R_1,...,R_N)$  taking values on $1..N$ and depending on $W$, such that, for each $1\leq i\leq N$, %, letting $F_i$ denote the number of occurrences of $i$ in $R$,
one has:
$\expc[\sum_{j=1}^N  1_{R^{(j)}=i}\,|W]=N\cdot \widehat W^{(i)}%/(\sum_{j=1}^N W_j)
%\left[\sum_{j=1}^N 1\{R^{(j)}=k\}\right]=N\cdot W(k)\,.
$.
In other words,      each index $i\in 1..N$   on average   is selected in $R$ a number of times  proportional to its  weight in $W$. We shall write $R(W)$ to indicate that $R$ depends on a given weight vector $W$.  Various   resampling schemes have been proposed in the literature, among which the simplest is perhaps \emph{multinomial resampling}; see e.g.  \cite[Ch.9]{SMC} and references therein.  Algorithm \ref{alg:PF} presents a generic  PF algorithm. Resampling here takes place at step 4: its purpose is to give more importance to particles with higher weight, when extracting the next generation of $N$ particles, while discarding particles with lower weight.

%{eq:filtering}
The justification and usefulness of this algorithm is that, under mild assumptions  \cite{SMC},   for any measurable function $h$ defined on $\X$, expectation under $\phi_t$, the filtering distribution on $\X$ at time $t$, in the limit can be expressed a weighted sum with weights $\widehat W^{(j)}_t$:
\vspace*{-0.3cm}
{\small
\begin{equation}\label{eq:convPF}
	\sum_{j=1}^N \widehat W^{(j)}_t\cdot h(X^{(j)}_t)\, \longrightarrow\,\expc_{\phi_t}[h] \text{\ \ \ \  a.s. as } N\longrightarrow +\infty\,.
	\vspace*{-0.3cm}
\end{equation}
}\noindent
%where $\phi_t$ is the filtering distribution  at time $t$ on $\X$.
The practical implication here is that we can estimate quite effectively    the expectations   involved in \eqref{eq:filtering}, for $\phi_t=\phi_{S,t}$, as weighted sums like in \eqref{eq:convPF}. %, even if the required sample size  for accurate estimation is generally quite large.
%The required sample size can be quite large also exponential when considering  Importance Sampling \cite{Chatterjee}, \lui{aggiunto rif. Chatterjee } but this is generally balanced out by greater accuracy, as can be seen also in Table \ref{tab:table1} in Appendix \ref{app:air}.
Note that in the above consistency   statement  $t$ is held fixed --- it is one of the parameter of the FK model --- while the number of particles $N$ tends to $+\infty$.
% (note that here $t$ is held fixed, while $N$ varies).
%Perhaps the simplest is letting $R$ be $N$ i.i.d. random variables each distributed according to $\widehat W$: this is known as  . We refer the reader to the specialized literature on PF for details and efficient implementation methods,
%
%Finally, for the sake of uniform notation we  convene that, at the first step ($k=1$),  $K_1(\xi_{-1}):=\mu^1$.

\begin{algorithm}[t]{\small
	\caption{A generic PF algorithm}
	\label{alg:PF}
	{\small
		\begin{algorithmic}[1]
			%\Procedure{Roy}{$a,b$}
			\Statex \textbf{Input}:  {$\FC=(\X,t,\mu^1,\{K_k\}_{k=2}^t,\{G_k\}_{k=1}^t)$, a FK model; $N\geq 1$, no. of particles.}
			\Statex \textbf{Output}:{$X^{1:N}_{t}\in \X^N$, $W^{1:N}_t\in \reals^{+N}$.}
			\State  $X^{(j)}_1 \sim \mu^1 $\Comment{state initialisation}\tikzmark{top}
			\State $ W^{(j)}_{1} := G_1(X^{(j)}_{1})$\ \ \tikzmark{bottom}\tikzmark{right}\Comment{weight initialisation}
			%\State $\hat w^{(1)}_{j} = w^{(1)}_{j}/W_1$ %($S_1=\sum_{k=1}^N w^{(1)}_k$) for $j=1,...,N$
			\For{ $k=2,...,t$ }
			\State $r_{1:N}\sim R(W^{1:N}_{k-1})$\Comment{resampling}%\tikzmark{topT}
			\State  $X^{(j)}_k \sim K_k(X^{(r_j)}_{k-1})$\ \ \tikzmark{topT}\tikzmark{rightT}\Comment{state update}
			\State $W^{(j)}_{k}:=G_{k}(X^{(j)}_{k})$ \tikzmark{bottomT}\Comment{weight update}
			%\State $\hat w^{(k)}_{j}=w^{(k)}_{j}/W_{k}$
			\EndFor
			\State\Return $(X_t ,W_t )$%$(X^{1:N}_t,W^{1:N}_t)$
		\end{algorithmic}
		\label{key}		\AddNote{top}{bottom}{right}{\ \    ($j=1,...,N$)}
		\AddNote{topT}{bottomT}{rightT}{\ \  ($j=1,...,N$)}
		%\caption{Bla}
}}%\label{alg:PF}
\vspace*{-0.4cm}
\end{algorithm}

\ifmai
\section{Implementation and experimental validation}\label{sec:experiments}
\paragraph{Implementation}
\vspace*{-0.08cm}
The PPG model is naturally amenable to a vectorized implementation  of PF that leverages the fine-grained, SIMD parallelism existing at the level of particles. At every iteration,  the state of the $N$ particles, $\omega^N=(\omega_1,...,\omega_N)$  with $\omega_i=(v_i,z_i)\in \ereals^{m+1}$, will be stored using a pair of arrays $(V,Z)$  of   shape ${N\times m}$  and   ${N\times 1}$, respectively. The weight vector is stored using another array $W$ of shape ${N\times 1}$. We rely on  {vectorization} of operations: for a  function $f:\ereals^k\rightarrow \ereals$ and a $N\times k$ array $U$,  $f(U)$ will denote the $N\times 1$  array obtained by applying $f$   to each row of $U$. In particular, we denote by $(Z=s)$ (for any $s\in \mathbb{N}$) the $N\times 1$ array obtained applying   element-wise the indicator function $1_{\{s\}}$  to $Z$ element-wise, and by $\varphi(V)$ the $N\times 1$ array obtained by applying the predicate $\varphi$ to $V$. For $U$   a $N\times k$   array and $W$ a $N\times 1$ array, $U\ast W$ denotes the $N\times k$   array obtained by multiplying the   $j$th row of $U$ by the $j$th element of $W$, for $j=1,...,N$: when $W$ is a $0/1$ vector, this is an instance of \emph{boolean masking}. Abstracting the vectorization primitives of modern CPUs and programming languages, we model the assignments  of  a vector to an array variable    as a single instruction,
%\footnote{Actual execution  as a single machine instruction will also depend  on the size of the involved arrays and on   specifics of the hardware.},
written $U:=Z$. The usual rules for broadcasting scalars to vectors apply, so e.g. $V:=S$ for $S\in \ereals$ means filling $V$ with $S$.  Likewise, for $\zeta$ a parametric distribution, $U\sim \zeta(V)$ means sampling $N$ times independently from   $\zeta(v_1),...,\zeta(v_N)$, and assigning the resulting matrix to $U$: this too counts as a single instruction.

Based on the above idealized model of vectorized computation,  we present VPF, a vectorized  version of the  PF algorithm for PPGs, as Algorithm \ref{alg:VPF}.   Here it is assumed that $\St\subseteq \mathbb{N}$, while $\psco(s)=\gamma_s$. On line 4, $\text{Resampling}(\cdot)$ denotes the result of applying a generic resampling algorithm based on weights $W$ to the current particles' state, represented by the   pair of vectors $(V,Z)$. With respect to the generic PF Algorithm \ref{alg:PF}, here in  the returned output, $(V,Z)$ corresponds to $X_t$  and $W$ to $W_t$.
%boolean masking and multiple   {\verb"tf.where"} expressions. The main iteration is implemented via a single {\verb"tf.while_loop"} command; see \cite{github} for additional details.

\begin{algorithm}[t]{\small
	%\caption{PF algorithm for PPGs}
	{\small
		\begin{algorithmic}[1]
			%\Procedure{Roy}{$a,b$}
			\Statex \textbf{Input}:  {$\ppg=(\St,E,\nil,\psco)$, a PPG; $S\in \St$, initial program checkpoint; $t\geq 1$, time horizon; $N\geq 1$, no. of particles.}
			\Statex \textbf{Output}:{ $V \in \ereals^{m\times N}$, $Z,W\in \ereals^{1\times N}$.}
			\State  $V:=S $\ ;\  $Z:=S $\Comment{state initialisation}
			%\State  $Z:=S $
			\State $ W := \gamma_S(Z)$\Comment{weight  initialisation}
			%\State $\hat w^{(1)}_{j} = w^{(1)}_{j}/W_1$ %($S_1=\sum_{k=1}^N w^{(1)}_k$) for $j=1,...,N$
			\For{ $t-1\text{ \textbf{times}}$}
			\State $(V,Z):=\mathrm{Resampling}((V,Z),W)$\Comment{resampling}%\tikzmark{topT}
			\For{ $(s,\varphi,\zeta,s')\in E$ }
			\State $M_{s,\varphi}:=\varphi(V)\ast (Z=s)$\Comment{mask computation}
			\EndFor
			\State  $V \sim \sum_{(s,\varphi,\zeta,s')\in E}\, \zeta(V)\ast M_{s,\varphi}$\ \ ;\ \  $Z := \sum_{(s,\varphi,\zeta,s')\in E} \, \,s'\cdot M_{s,\varphi}$\Comment{state update}
			%\State  $Z := \sum_{(s,\varphi,\zeta,s')\in E} \, \,s'\cdot M_{s,\varphi}$\Comment{checkpoint update}
			\State  $W := \sum_{ s \in \St}\, \gamma_s(V)\ast (Z=s)$\Comment{weight update}
			%\State $\hat w^{(k)}_{j}=w^{(k)}_{j}/W_{k}$
			\EndFor
			\State\Return $(V,Z,W)$
		\end{algorithmic}
		%\vspace*{-0.3cm}
		%\label{alg:VPF}
	}	%\label{alg:VPF}
	\caption{VPF, a Vectorized PF algorithm  for PPGs.}\label{alg:VPF}}
\vspace*{-0.5cm}
\end{algorithm}
%\vspace*{-0.7cm}
%\lui{il pedice $k$ ad $r$ il Alg. 1 è proprio necessario?}

%\vspace*{-1cm}
\paragraph{Experimental validation}
We   illustrate some   experimental results obtained with a proof-of-concept  TensorFlow-based \cite{TF} implementation of Algorithm \ref{alg:VPF} (\TSIpf).  We have considered a number of challenging probabilistic programs that feature conditioning inside loops.   For all these programs, we will estimate $\sem S f$, for given functions $f$,  relying on the bounds provided by Theorem \ref{th:filtlift}  in terms of expectations w.r.t. filtering distributions. Such expectations will be estimated via \TSIpf.
%Overall, what follows should be interpreted just as an experimental validation of our approach.  Nevertheless, the results are quite encouraging, as discussed below.
We also compare \TSIpf\ with two state-of-the-art PPLs, webPPL \cite{webppl} and CorePPL \cite{CorePPL}.
%webPPL is  a popular PPL supporting several inference algorithms, including SMC, where resampling is handled via continuation passing.   We have chosen to   consider CorePPL  as it supports a very efficient implementation of PF.
In \cite{CorePPL},  a comparison of CorePPL with webPPL, Pyro \cite{Pyro} and other PPLs in terms of performance shows the superiority of CorePPL SMC-based inference across a number of benchmarks. %\footnote{Direct  compilation of CorePPL to GPU  via the  intermediate-level format   RootPPL     is also supported. However, the results we have obtained with RootPPL are generally worse in terms of  execution time, and not presented here. Our PC configuration is as follows. OS: Windows 10; CPU: 2.8 GHz Intel Core i7; GPU: Nvidia T500, driver v. 522.06; TF: v. 2.10.1; CUDA Toolkit v. 11.8; cuDNN SDK v. 8.6.0.}.

 At least for $N\geq 10^5$,   the tools tend generally to  return   similar estimates of the expected value, which we take    as an empirical evidence of accuracy. Additional insight into accuracy is obtained by directly comparing the results of VPF with those of webPPL-rejection (when available), which is an exact inference algorithm. The expected values estimated by webPPL-rejection  are consistently in line to those of VPF.
In terms of performance, a graphical representation of our results is provided is provided in Figure \ref{fig:scatterplot}, with scatterplots showing the ratio of execution times $(time_{\mathrm{other-tool}} / time_{\mathrm{VPF}})$ on a log scale.
In the case of WebPPL, nearly all data points lie above the x-axis, indicating superiority of VPF. In the case of CorePPL, for $N=10^5$ the data points are  quite uniformly distributed  across the x-axis, indicating basically a tie. For $N=10^6$, we have a majority of points above  the x-axis, indicating again superiority of VPF by \emph{orders of magnitude}; additional details can be found in Appendix \ref{app:exp}.
\fi
%\mik{Rivedere questa sezione.}
% of   \TSIpf\ over the other tools.
%%Overall, we take this as an evidence of the higher scalability of   \TSIpf\ over the other tools (additional details in the caption of Table \ref{tab:table1}).
%A closer look in the $N=10^6$  case reveals that the only programs where CorePPL beats VPF
%are RW1 and ZC. This is most likely due to the low probability of conditioning  in these programs; for instance in  RW1  just a single final conditioning is performed. As in CorePPL   resampling  is only performed following a conditioning, this may   explain its lower execution times in these cases.

\begin{algorithm}[t]{\small
%\caption{PF algorithm for PPGs}
{\small
	\begin{algorithmic}[1]
		%\Procedure{Roy}{$a,b$}
		\Statex \textbf{Input}:  {$\ppg=(\St,E,\nil,\psco)$, a PPG; $S\in \St$, initial pr. checkpoint; $t\geq 1$, time horizon; $N\geq 1$, no. of particles.}
		\Statex \textbf{Output}:{ $V \in \ereals^{m\times N}$, $Z,W\in \ereals^{1\times N}$.}
		\State  $V:=S $\ ;\  $Z:=S $\Comment{state initialisation}
		%\State  $Z:=S $
		\State $ W := \gamma_S(Z)$\Comment{weight  initialisation}
		%\State $\hat w^{(1)}_{j} = w^{(1)}_{j}/W_1$ %($S_1=\sum_{k=1}^N w^{(1)}_k$) for $j=1,...,N$
		\For{ $t-1\text{ \textbf{times}}$}
		\State $(V,Z):=\mathrm{Resampling}((V,Z),W)$\Comment{resampling}%\tikzmark{topT}
		\For{ $(s,\varphi,\zeta,s')\in E$ }
		\State $M_{s,\varphi}:=\varphi(V)\ast (Z=s)$\Comment{mask computation}
		\EndFor
		\State  $V \sim \sum_{(s,\varphi,\zeta,s')\in E}\, \zeta(V)\ast M_{s,\varphi}$\ \ ;\ \  $Z := \sum_{(s,\varphi,\zeta,s')\in E} \, \,s'\cdot M_{s,\varphi}$\Comment{state update}
		%\State  $Z := \sum_{(s,\varphi,\zeta,s')\in E} \, \,s'\cdot M_{s,\varphi}$\Comment{checkpoint update}
		\State  $W := \sum_{ s \in \St}\, \gamma_s(V)\ast (Z=s)$\Comment{weight update}
		%\State $\hat w^{(k)}_{j}=w^{(k)}_{j}/W_{k}$
		\EndFor
		\State\Return $(V,Z,W)$
	\end{algorithmic}
%\vspace*{-0.3cm}
	%\label{alg:VPF}
}	%\label{alg:VPF}
\caption{VPF, a Vectorized PF algorithm  for PPGs.}\label{alg:VPF}}
\vspace*{-0.1cm}
\end{algorithm}
%\vspace*{-0.7cm}
%\lui{il pedice $k$ ad $r$ il Alg. 1 è proprio necessario?}
\section{Implementation and experimental validation}\label{sec:experiments}
\subsection{Implementation}
\vspace*{-0.08cm}
The PPG model is naturally amenable to a vectorized implementation  of PF that leverages the fine-grained, SIMD parallelism existing at the level of particles. At every iteration,  the state of the $N$ particles, $\omega^N=(\omega_1,...,\omega_N)$  with $\omega_i=(v_i,z_i)\in \ereals^{m+1}$, will be stored using a pair of arrays $(V,Z)$  of   shape ${N\times m}$  and   ${N\times 1}$, respectively. The weight vector is stored using another array $W$ of shape ${N\times 1}$. We rely on  {vectorization} of operations: for a  function $f:\ereals^k\rightarrow \ereals$ and a $N\times k$ array $U$,  $f(U)$ will denote the $N\times 1$  array obtained by applying $f$   to each row of $U$. In particular, we denote by $(Z=s)$ (for any $s\in \mathbb{N}$) the $N\times 1$ array obtained applying   element-wise the indicator function $1_{\{s\}}$  to $Z$ element-wise, and by $\varphi(V)$ the $N\times 1$ array obtained by applying the predicate $\varphi$ to $V$  to the row-wise. For $U$   a $N\times k$   array and $W$ a $N\times 1$ array, $U\ast W$ denotes the $N\times k$   array obtained by multiplying the $j$th row of $U$ by the $j$th element of $W$, for $j=1,...,N$: when $W$ is a $0/1$ vector, this is an instance of \emph{boolean masking}. Abstracting the vectorization primitives of modern CPUs and programming languages, we model the assignments  of  a vector to an array variable    as a single instruction,
%\footnote{Actual execution  as a single machine instruction will also depend  on the size of the involved arrays and on   specifics of the hardware.},
written $U:=Z$. The usual rules for broadcasting scalars to vectors apply, so e.g. $V:=S$ for $S\in \ereals$ means filling $V$ with $S$.  Likewise, for $\zeta$ a parametric distribution, $U\sim \zeta(V)$ means sampling $N$ times independently from   $\zeta(v_1),...,\zeta(v_N)$, and assigning the resulting matrix to $U$: this too counts as a single instruction.

Based on the above idealized model of vectorized computation,  we present VPF, a vectorized  version of the  PF algorithm for PPGs, as Algorithm \ref{alg:VPF}.   Here it is assumed that $\St\subseteq \mathbb{N}$, while $\psco(s)=\gamma_s$. On line 4, $\text{Resampling}(\cdot)$ denotes the result of applying a generic resampling algorithm based on weights $W$ to the current particles' state, represented by the   pair of vectors $(V,Z)$. With respect to the generic PF Algorithm \ref{alg:PF}, here in  the returned output, $(V,Z)$ corresponds to $X_t$  and $W$ to $W_t$. Note that there are no loops where the number of iterations depends on $N$;  the \textbf{for} loop in lines 5--7   only scans the transitions set $E$, whose size is independent of $N$. Line 8 is just a vectorized implementation of sampling from the Markov kernel function in \eqref{eq:FCM}. Line 9 is a vectorized implementation of the combined score function \eqref{eq:scoz}.
In the actual TensorFlow implementation, the sums in lines 8 and 9  are encoded via boolean masking and vectorized operations.
%boolean masking and multiple   {\verb"tf.where"} expressions. The main iteration is implemented via a single {\verb"tf.while_loop"} command; see \cite{github} for additional details
\vspace*{-.3cm}
%\vspace*{-0.5cm}
\subsection{Experimental validation}
We   illustrate some   experimental results obtained with a proof-of-concept  TensorFlow-based \cite{TF} implementation of Algorithm \ref{alg:VPF}. We still refer to this implementation as \TSIpf.  We have considered a number of challenging probabilistic programs that feature conditioning inside loops.   For all these programs, we will estimate $\sem S f$, for given functions $f$,  relying on the bounds provided by Theorem \ref{th:filtlift}  in terms of expectations w.r.t. filtering distributions. Such expectations will be estimated via \TSIpf.
%Overall, what follows should be interpreted just as an experimental validation of our approach.  Nevertheless, the results are quite encouraging, as discussed below.
We also compare \TSIpf\ with two state-of-the-art PPLs, webPPL \cite{webppl} and CorePPL \cite{CorePPL}. webPPL is  a popular PPL supporting several inference algorithms, including SMC, where resampling is handled via continuation passing.   We have chosen to   consider CorePPL by Lunden et al. as it supports a very efficient implementation of PF: in \cite{CorePPL},  a comparison of CorePPL with webPPL, Pyro \cite{Pyro} and other PPLs in terms of performance shows the superiority of CorePPL SMC-based inference across a number of benchmarks.
%; CorePPL's implementation  is based on a compilation into an intermediate format, conceptually similar to our PPGs.
%\footnote{Direct  compilation of CorePPL to GPU  via the  intermediate-level format   RootPPL     is also supported. However, the results we have obtained with RootPPL are generally worse in terms of  execution time, and not presented here. Our PC configuration is as follows. OS: Windows 10; CPU: 2.8 GHz Intel Core i7; GPU: Nvidia T500, driver v. 522.06; TF: v. 2.10.1; CUDA Toolkit v. 11.8; cuDNN SDK v. 8.6.0.}.

\vspace*{-.3cm}
%\mik{Attenzione! Cambiato R2 in ZC e R3 in R2.}
\paragraph{Models} For our experiments we have considered the following  programs:
 \emph{Aircraft tracking} (AT, \cite{WuEtAl}), \emph{Drunk man and mouse} (DMM, Example \ref{ex:dm1}), \emph{Hare and tortoise} (HT, e.g. \cite{Bagnall}), \emph{Bounded retransmission protocol} (BRP, \cite{5}), \emph{Non-i.i.d. loops} (NIID, e.g. \cite{5}), the \emph{ZeroConf} protocol (ZC, \cite{2}), and two variations of \textit{Random Walks},  RW1 (\cite{VMCAI24}, Example 2) and RW2 in the following. In particular, AT is a model where a single aircraft is tracked in a 2D space using noisy measurements from six radars.
  %For  DMM, see Example \ref{ex:dm1}.
  HT simulates a race between a hare and a tortoise on a discrete line. BRP models a scenario where multiple packets are transmitted over a lossy channel.    NIID   describes a process that keeps tossing two fair coins until both show tails. ZC is an idealized version of the network connection protocol by the same name.  RW1, RW2 are random walks with Gaussian steps. The pseudo-code of these models  is reported in \cite{BC25}.% Appendix \ref{app:models}.\
These programs feature conditioning/scoring inside loops.
In particular, DMM, HT and NIID feature unbounded loops: for these three programs, in the case of \TSIpf\ we have truncated the execution after   $k=1000,100,100$ iterations, respectively, and set the time parameter $t$ of Theorem \ref{th:filtlift} accordingly, which allows us to deduce bounds on the value of $\sem S f$ (for the precise definition of $f$ in each case, see  \cite{BC25}).  For the other tools, we just consider the   truncated estimate returned  at the end of   $k$ iterations.  AT, BRP, ZC,  RW1 and RW2  feature bounded loops, but are nevertheless quite challenging. In particular,  AT   features multiple conditioning inside a for-loop,  sampling from a mix of continuous and discrete distributions, and noisy observations.
Below, we discuss the obtained experimental results in terms of accuracy and performance.
%A   description  of these programs, together with  further details on the experimental set up, can be found in \iffull Appendix \ref{app:air} and \cite{github};\else \cite{BC25,github}.\fi
% code available from \cite{github}. %Code available from \cite{github}.
%
%\ Table \ref{tab:table1}  summarizes the obtained experimental results in terms, which we comment in the following.

\vspace*{-.4cm}
\paragraph{Accuracy}
%We report in Table \ref{tab:table1} (Appendix) the execution time, the estimated expected value and the Effective Sample Size (ESS, a measure of diversity of particles, the higher the better; \iffull see Appendix \ref{app:air}\else see \cite{BC25}\fi) for \TSIpf, CorePPL and webPPL,
We have compared \TSIpf, CorePPL and webPPL across the above mentioned examples for different values of $N$, the number of particles
(details  in \cite[Table 1]{BC25}).
%Table \ref{tab:\TSIpf, CorePPL and webPPL, table1} (Appendix)
At least for $N\geq 10^5$,   the tools tend generally to  return very  similar estimates of the expected value, which we take    as an empirical evidence of accuracy. Additional insight into accuracy is obtained by directly comparing the results of VPF with those of webPPL-rejection (when available), which is an exact inference algorithm: the expected values estimated by webPPL-rejection  are consistently in line to those of VPF.  We have also considered Effective Sample Size (ESS), a measure of   diversity of the sample,  the higher the better \cite{RobertESS}. In terms of ESS, the difference across the tools is significant: with one exception (program RW1),  VPF  yields ESS that are higher  or comparable to those of the other tools. We refer the reader to \cite{BC25} for additional explanation, in particular as to the significance of the mentioned exception.
%For the other models, we provide the lower bound of the interval.
%In particular, the effective sample size (ESS) quantifies diversity of the particles (the higher the better \cite{RobertESS}). Essentially, an high ESS indicates that the samples are rather representative of the target distribution, since they are close to the behavior of independent samples. In fact, there exist asymptotic guarantees of correctness for parametric estimates based on independent samples; the same is not true when the samples are correlated.
%
%The four analyzed tools generate estimates that are very close to each other, therefore they are quite similar in terms of accuracy.
%
\vspace*{-.4cm}
\paragraph{Performance} For larger values of $N$   \TSIpf\  generally outperforms   the other   considered tools   in terms of execution time. The difference is especially noticeable  for $N=10^6$.
Figure \ref{fig:scatterplot} provides a graphical comparison,  with scatterplots showing the ratio of execution times $(time_{\mathrm{other-tool}} / time_{\mathrm{VPF}})$ on a log scale, across the different examples (actual data points in  \cite[Table 1]{BC25}).
In the case of WebPPL, nearly all data points lie above the x-axis, indicating superiority of VPF. In the case of CorePPL, for $N=10^5$ the data points are  quite uniformly distributed  across the x-axis, indicating basically a tie. For $N=10^6$, we have a majority of points above  the x-axis, indicating again superiority of VPF, sometimes by orders of magnitude.
\newcommand{\scalefactor}{8.0}
\begin{figure}[t]
\centering
\includegraphics[width=4.4cm,height=3.45cm]{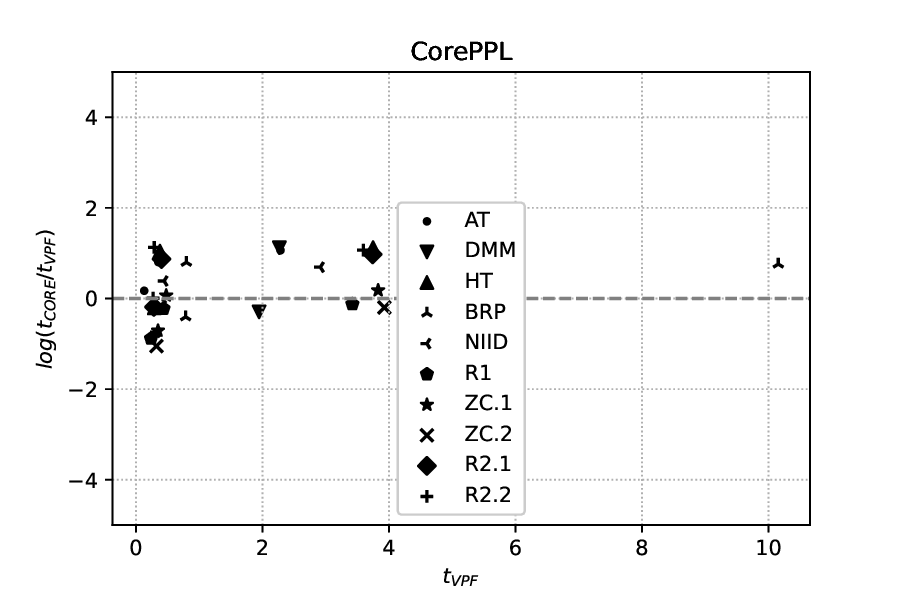}\hspace*{-0.5cm}
\includegraphics[width=4.4cm,height=3.45cm]{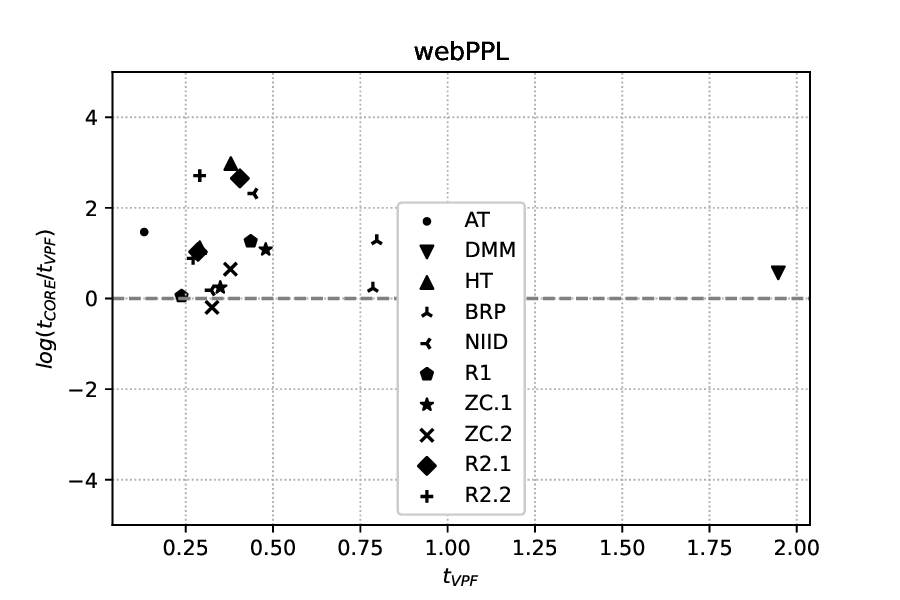}\hspace*{-0.4cm}%\\
\includegraphics[width=4.4cm,height=3.45cm]{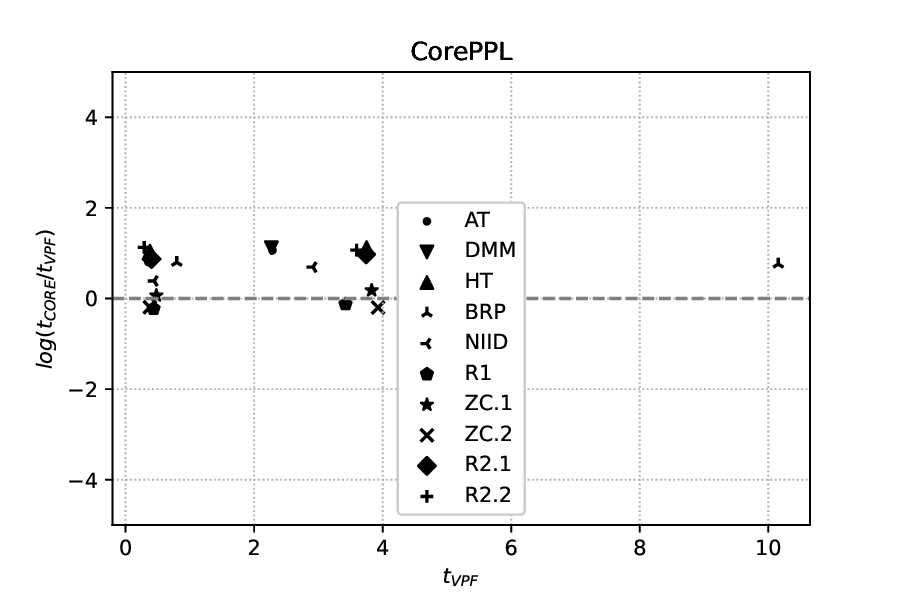}\hspace*{-0.5cm}
\includegraphics[width=4.4cm,height=3.45cm]{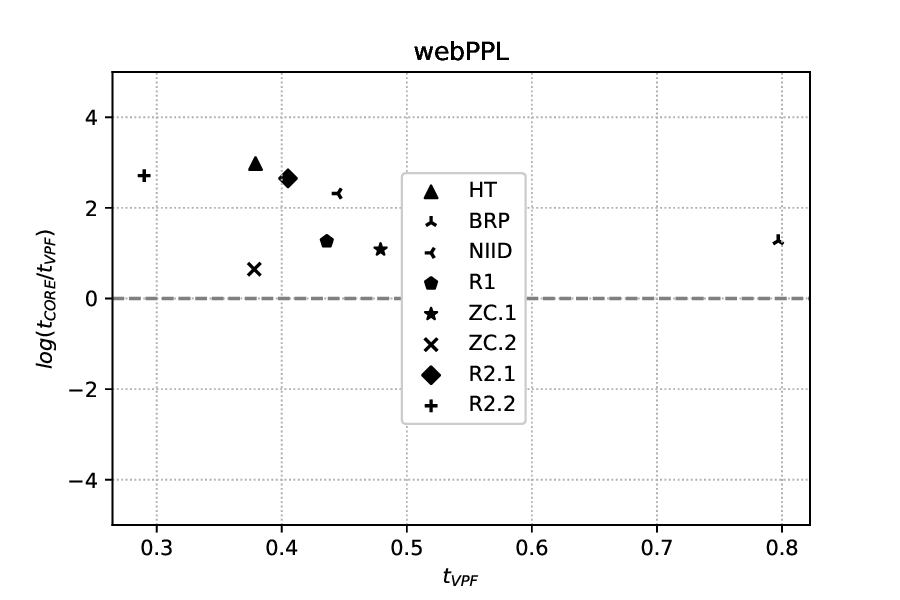}
	\vspace{-0.7cm}
\caption{\footnotesize%\scriptsize
	For $N=10^5,10^6$, scatterplots of the log-ratios of execution times, $\log_{10}(\mathrm{time}_{{tool}}/\mathrm{time}_{\mathrm{VPF}})$, based on   data points in \cite[Table 1]{BC25}. From left to right: $N=10^5$, $tool=$ CorePPL;  $N=10^5$, $tool=$ WebPPL-smc;  $N=10^6$, $tool=$ CorePPL;  $N=10^6$, $tool=$ WebPPL-smc.
	For $N=10^6$,  the vast majority of the data points lie above the x-axis, indicating superior performance of VPF across different examples.		
}
\label{fig:scatterplot}
\vspace{-0.3cm}
\end{figure}

\ifmai
\mik{Rivedere questa sezione.}
% of   \TSIpf\ over the other tools.
 %%Overall, we take this as an evidence of the higher scalability of   \TSIpf\ over the other tools (additional details in the caption of Table \ref{tab:table1}).
A closer look in the $N=10^6$  case reveals that the only programs where CorePPL beats VPF
are RW1 and ZC. This is most likely due to the low probability of conditioning  in these programs; for instance in  RW1  just a single final conditioning is performed. As in CorePPL   resampling  is only performed following a conditioning, this may   explain its lower execution times in these cases.  To further investigate this issue, we consider   RW2, where the probability of conditioning is governed by a parameter $\lambda\in [0,1]$, and run it for different values of $\lambda$.
%, where $\lambda$ represents the probability of performing a conditioning step.
%We focus on R3, as termination and resampling probabilities are independent.
The obtained results are showed in Figure \ref{fig:time_ev}. We   observe that for both CorePPL and WebPPL execution time tends to increase  as the probability  $\lambda$ of conditioning  increases; on the contrary,  the execution time of VPF appears to be insensitive to  $\lambda$. This suggests that VPF has a definite advantage over tools with explicit resample,  on models with heavy conditioning.
\fi
%Figure \ref{fig:time_ev} also shows that, for both CoRePPL and VPF, the the specific boolean  of the observe condition has no a significant effect on execution time.% and does not constitute a computational bottleneck.

%The difference between CorePPL and VPF will be further analyzed  in the next paragraph   in terms of scalability.

%\mik{Ricontrollare. Rimuovere la figure e inserire il plot nel running text, a destra (con wrapfigure o simili).}
%\paragraph{Scalability}   The plot on the right  shows the behaviour the \emph{average unit cost} of VPF, CorePPL and WebPPL across all the models we analyzed for $N=10^3,...,10^6$ on a log-scale. Here, for each $N$ the average unit cost is $(t_1+t_2+..+t_k)/(N\cdot k)$, with $t_i$ the execution time of the $i-th$ example.
%Consistently with Figure \ref{fig:time_ev}, we can observe that the cost of VPF decreases as the number of samples increases, whereas the cost of the other tools remains   constant or increases (webPPL).
%
%\begin{figure}[t]
%		\begin{center}
%			\includegraphics[width=5.0cm,height=3.9cm]{TT.eps}
%		\end{center}
%	\vspace{-0.5cm}
%	\caption{{ Average unit cost of VPF, CorePPL and webPPL as the number of samples increases.\textbf{ Left:} numerical values of the unit cost for the considered tools. \textbf{Right:} their graphical representation on a logarithmic scale.}}
%	\label{fig:tt}
%\end{figure}

\ifmai
\paragraph{Scalability}
\begin{wrapfigure}{r}{4.0cm}
	\vspace*{-1.2cm}
	\includegraphics[width=4.0cm,height=2.9cm]{TT.eps}
\end{wrapfigure}
The plot on the right shows the behaviour the \emph{average unit cost (per particle)} of VPF, CorePPL and WebPPL across all the models we analyzed for $N=10^3,...,10^6$ on a log-scale. Here, for each $N$ the average unit cost (in seconds) is $(t_1+t_2+..+t_k)/(N\cdot k)$, with $t_i$ the execution time of the $i$-th example. Consistently with Figure \ref{fig:time_ev}, we can observe that the cost of VPF decreases as the number of samples increases, whereas the cost of the other tools remains constant or increases (webPPL).
\fi
%most cases and its execution times are always the lowest when the number of particles is high .

%In particular, the effective sample size quantifies diversity of the particles, the higher the better \cite{RobertESS}) as the number of particles increases, for \TSIpf, CorePPL, RootPPL and webPPL.

\ifmai
\section{Conclusion}\label{sec:concl}%\begin{minipage}{\textwidth}
\vspace*{-.2cm}
We study   correct and efficient implementations of Sequential Monte Carlo  inference algorithms for   universal probabilistic programs. We offer a clean trace-based operational semantics for PPGs,
a finite approximation theorem and  consistency of the PF algorithm via a connection to FK models.
 Experiments conducted with VPF, a vectorized version of PF tailored to PPGs, show  very promising results.
\fi

\section{Conclusion}\label{sec:concl}
%We have proposed a framework for inference in universal probabilistic programs that combines a rigorous  trace-based operational semantics with practical algorithmic realizability.
%The approach builds on a structured operational model and connects to classical tools in stochastic process theory (FK models), leading to an inference method that is both theoretically sound and efficiently implementable.
We summarize the main insights of our approach and possible future directions below.
\vspace*{-.3cm}
\paragraph{Structured semantics via PPGs (Sections \ref{sec:PP} and \ref{sec:obs})}
Our framework is centered around   an automata-theoretic description format for programs, {Probabilistic Program Graphs} (PPGs). %This automata-theoretic formalism captures the control-flow and stochastic behavior of probabilistic programs in a way that is both mathematically precise and amenable to implementation.
In PPGs,   transitions encode  sampling behavior, while nodes represent conditioning checkpoints via score functions. This structure supports a rigorous infinite-trace semantics  and facilitates the alignment of computations, a feature that becomes crucial for vectorized implementations.

\vspace*{-.3cm}
\paragraph{Approximation via Feynman-Kac models (Sections \ref{sec:FA} and \ref{sec:MC})}
The main theoretical contribution is a novel connection between the infinite-trace semantics of PPGs and  {Feynman-Kac} (FK) models, a standard tool in the analysis of state-based stochastic processes.  The expected values of a broad class of \emph{prefix-closed} functions over infinite traces can be expressed in terms of finite, truncated computations. In particular, Theorem~\ref{th:approx} establishes that these expectations can be bounded by quantities defined over finite-length traces, making inference tractable in the presence of unbounded loops and conditioning. Theorem~\ref{th:tight} shows that these bounds converge to the exact value under mild assumptions. We then establish a connection with FK models: in particular Theorem~\ref{th:filtlift}  reformulates the approximation bounds in terms of the \emph{filtering} distributions of a  PPG-induced FK model —distributions that can be consistently and efficiently estimated via particle filtering (PF).

\vspace*{-.3cm}
\paragraph{Efficient and parallel inference (Section \ref{sec:experiments})}
A central insight of our approach is that the operational structure of PPGs enables a naturally parallelizable implementation of inference. Since all particles evolve synchronously (in lock-step) through the same control-flow graph (PPG), and conditioning  is applied via score functions in a uniform, aligned fashion, our particle filtering algorithm maps directly to SIMD-style vectorized execution. This design avoids the particle misalignment issues that affect continuation-based or functional semantics for PPLs. As a result, our operational model is easily mapped into  modern hardware architectures supporting data-level parallelism.
%
%\paragraph{Practical validation.}
Our vectorized implementation of a PPG-based particle filter, VPF, practically demonstrates the effectiveness of our approach. On challenging examples involving nested conditioning and unbounded loops, VPF matches or outperforms state-of-the-art probabilistic programming systems in both accuracy and runtime. %These results confirm that semantic rigor and algorithmic efficiency can go hand in hand.
\vspace*{-.3cm}
\paragraph{Future directions}
On the practical side, developing   compilers from high-level PPLs to PPGs, extending the framework to richer type systems and data structures is a natural next-step. Combining a sampling-based approach with symbolic or constraint-based reasoning techniques is a challenging theoretical direction.
%Overall, the integration of operational semantics, approximate inference, and parallel execution provides a robust foundation for both theoretical analysis and scalable implementation of probabilistic programming languages.

\ifmai
\newcommand{\scalefactor}{1.2}
\begin{figure}[t]
\centering
\includegraphics[width=4.31cm,height=3.45cm]{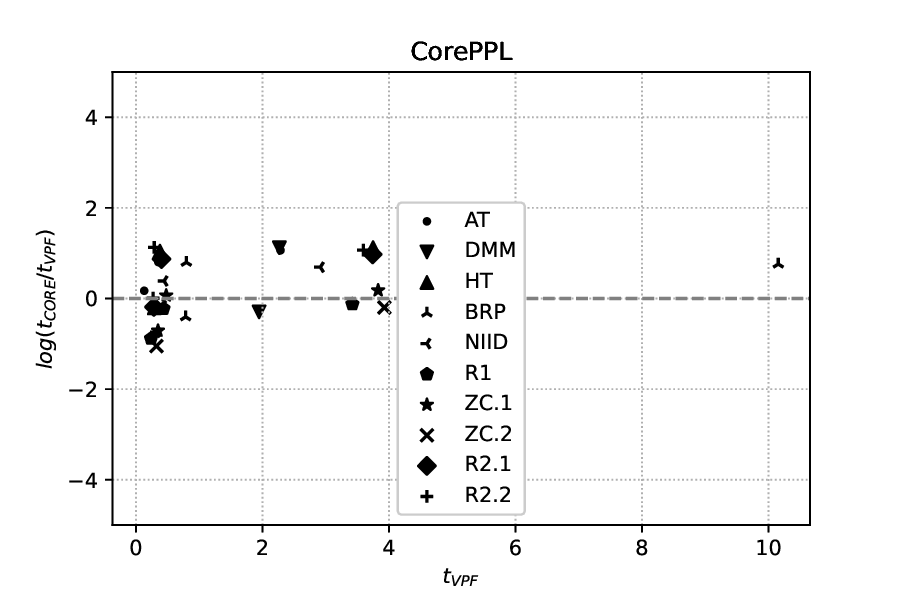}\hspace{-0.52cm}
\includegraphics[width=4.31cm,height=3.45cm]{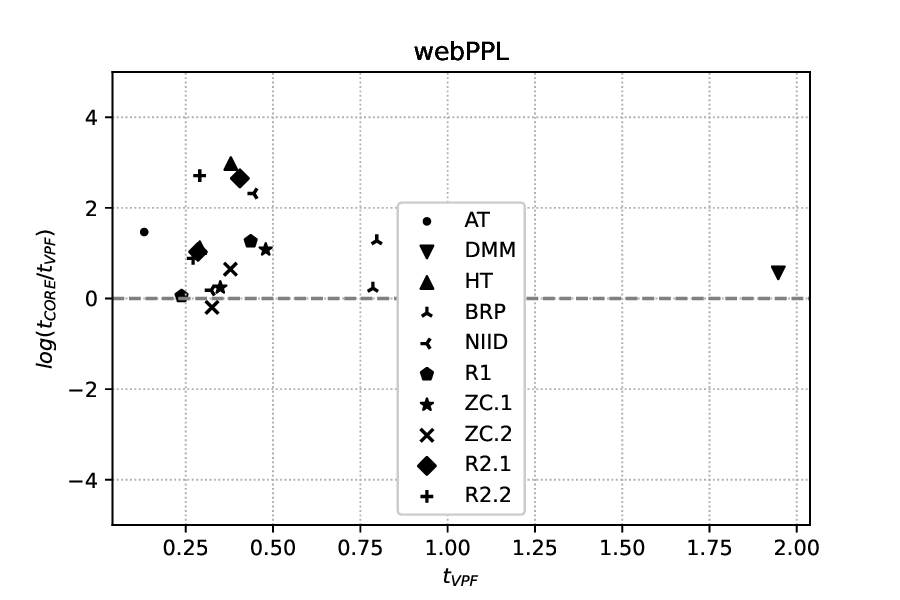}\hspace{-0.43cm}%\\
\includegraphics[width=4.31cm,height=3.45cm]{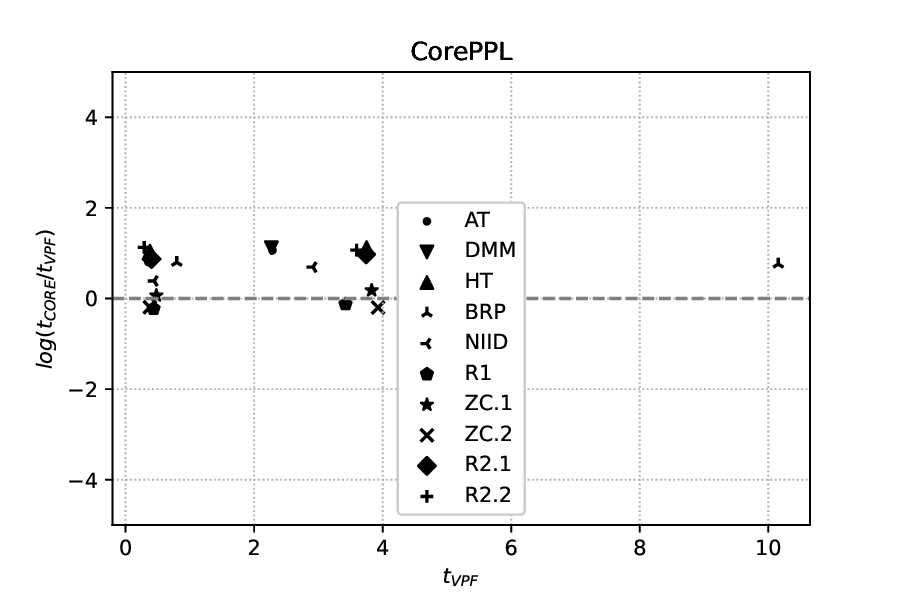}\hspace{-0.52cm}
\includegraphics[width=4.31cm,height=3.45cm]{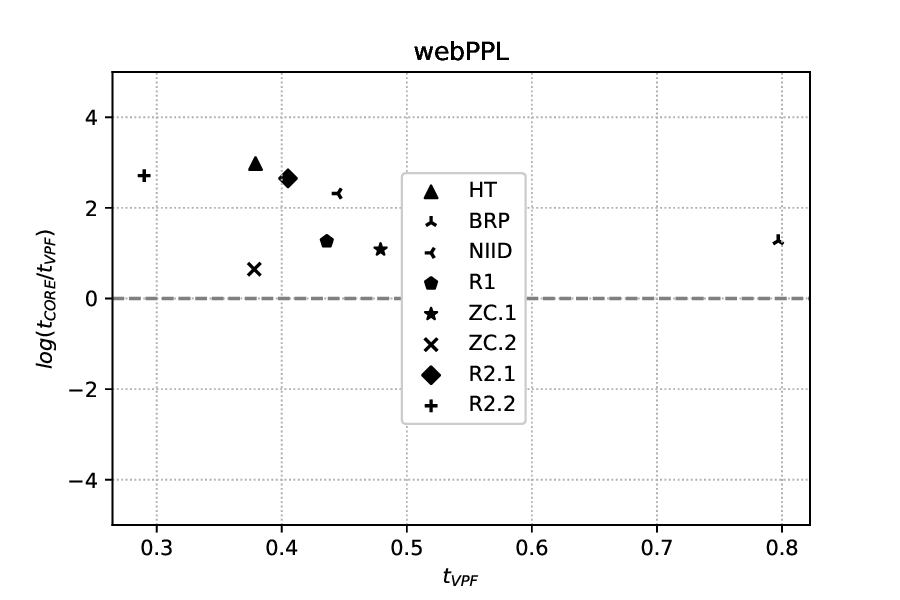}
	\vspace{-0.7cm}
\caption{\footnotesize
	For $N=10^5,10^6$, scatterplots of the log-ratios of execution times, $\log_{10}(\mathrm{time}_{{tool}}/\mathrm{time}_{\mathrm{VPF}})$, based on the data points of Table \ref{tab:table1}. From left to right: $N=10^5$, $tool=$ CorePPL;  $N=10^5$, $tool=$ WebPPL-smc;  $N=10^6$, $tool=$ CorePPL;  $N=10^6$, $tool=$ WebPPL-smc.
	For $N=10^6$,  the vast majority of the data points lie above the x-axis, indicating superior performance of VPF across different examples.		
}
\label{fig:scatterplot}
\vspace{-0.5cm}
\end{figure}

\begin{figure}[t]
	\centering
	\includegraphics[width=4.8cm,height=3.7cm]{timeT4.eps}
	\includegraphics[width=4.8cm,height=3.7cm]{timeT5.eps}
	\includegraphics[width=4.8cm,height=3.7cm]{timeT6.eps}
	\caption{\scriptsize Execution times (in seconds) for the RW2 program, as a function of the 	 probability $\lambda$  of conditioning on external data for $N=10^4$ (left), $N=10^5$ (center) and $N=10^6$ (right). webPPL   missing from the right-most plot due to time-out.
%when the observe statement condition is always satisfied (bottom) and when it is not for every input data (top).
%		Top and bottom plots exhibit the same behavior for CorePPL and VPF,
%		 suggesting that the strictness of the condition does not impact significantly their execution times. Moreover,
Execution times of VPF are basically insensitive to  $\lambda$.}
	\label{fig:time_ev}
\end{figure}
\fi

\newpage
\nocite{*}
\bibliographystyle{eptcs}
\bibliography{generic2}
\end{document}